\documentclass[11pt,preprint]{aastex}
\usepackage{epsfig}
\def\lsim{\raise0.3ex\hbox{$<$}\kern-0.75em{\lower0.65ex\hbox{$\sim$}}}
\def\gsim{\raise0.3ex\hbox{$>$}\kern-0.75em{\lower0.65ex\hbox{$\sim$}}}

\begin{document}

\title{Physical Properties of Dense Cores in the $\rho$ Ophiuchi Main
Cloud and A Significant Role of External Pressures 
in Clustered Star Formation}

\author{Hajime Maruta\altaffilmark{1}, 
Fumitaka Nakamura\altaffilmark{2,3}, Ryoichi Nishi\altaffilmark{1}, 
Norio Ikeda\altaffilmark{3}, Yoshimi Kitamura\altaffilmark{3}} 

\altaffiltext{1}{Department of Physics, Niigata University, 8050
Ikarashi-2, Niigata 950-2181, Japan}
\altaffiltext{2}{Astrophysics Lab., Faculty of Education, 
Niigata University, 8050 Ikarashi-2, Niigata 950-2181, 
Japan}
\altaffiltext{3}{Institute of Space and Astronautical Science, 
Japan Aerospace Exploration Agency, 3-1-1 Yoshinodai, Sagamihara, 
Kanagawa 229-8510, Japan}

%
%%%% limit of number of words is 250.
\begin{abstract}
Using the archive data of the H$^{13}$CO$^+$ ($J=1-0$) line
 emission taken with the Nobeyama 45 m radio telescope 
with a spatial resolution of $\sim 0.01$pc, 
we have identified 68 dense cores in the central dense region 
of the $\rho$ Ophiuchi main cloud.
The H$^{13}$CO$^+$ data also indicates that the fractional abundance 
of H$^{13}$CO$^+$ relative to H$_2$ is roughly inversely proportional 
to the square root of the H$_2$ column density 
with a mean of $1.72 \times 10^{-11}$.
The mean radius, FWHM line width, and LTE mass of the identified cores
are estimated to be 0.045 $\pm$ 0.011 pc, 0.49 $\pm$ 0.14 km s$^{-1}$, 
and 3.4 $\pm$ 3.6 $M_\odot$, respectively.
The majority of the identified cores have subsonic internal motions.
The virial ratio, the ratio of the virial mass to the LTE mass, 
tends to decrease with increasing the LTE mass and about
60 percent of the cores have virial ratios smaller than 2, 
indicating that these cores are not transient structures 
but self-gravitating.
The detailed virial analysis suggests that 
the surface pressure often dominates over the self-gravity and
thus plays a crucial role in regulating core formation and evolution.
By comparing the $\rho$ Oph cores with those in the Orion A molecular cloud
observed with the same telescope, 
we found that the statistical properties of the core physical quantities are 
similar between the two clouds
if the effect of the different spatial resolutions is corrected.
The line widths of the $\rho$ Oph cores appear to be
nearly independent of the core radii over the range of 0.01 $-$ 0.1 pc
and deviate upwards from the Heyer \& Brunt relation.
%%have significant excess from the Heyer \& Brunt relation.
This may be evidence that turbulent motions are driven 
by protostellar outflows in the cluster environment.
\end{abstract}


\section{Introduction}
\label{intro}

Observations of young stellar populations in the Galaxy 
have revealed that the majority of stars form in clusters
\citep{lada03,allen06}.
For example, \citet{lada91} performed an extensive near-infrared 
imaging survey of the central regions of L1630 in Orion B and found that 
about 60 $-$ 90 \% of all the young stellar populations associated with 
the molecular cloud form only in three rich embedded clusters. 
Using the 2MASS point source catalog, \citet{carpenter00} estimated
the fraction of young stellar populations contained in clusters 
to be 50 $-$ 100 \% for nearby cluster forming molecular clouds such as 
Perseus, Orion A, Orion B, and MonR2. 
Recent near-infrared surveys of young stellar populations using 
the Spitzer Space Telescope have confirmed that clustered star formation 
is the dominant mode 
of star formation in the Galaxy \citep{allen06,poulton08,romanzuniga08}.
However, previous studies of star formation have focused 
on star formation in relative isolation, for which the environmental
effects that play a crucial role in clustered star formation 
are thought to be minor or at least secondary.
Thus, how stars form in the cluster environment remains 
only poorly understood.

An important clue to understanding star formation in clusters 
has come from recent millimeter and submillimeter 
observations of the nearby cluster-forming region, 
the $\rho$ Ophiuchi molecular cloud (L1688), 
which uncovered  dense cores with a mass spectrum
that resembles the Salpeter IMF \citep{motte98,johnstone00, stanke06}.
Similar studies in other star forming regions have 
confirmed that such a resemblance between the core mass spectrum and the
stellar IMF seems to be common in nearby star forming regions
\citep{testi98, motte01, reid06, ikeda07, ikeda09}.
These observations suggest that the stellar IMF may be determined 
to a large extent by the core mass distribution
\citep[e.g.,][]{andre07},
although mass accretion from ambient gas onto forming stars may
sometimes plays an important role in determining the final masses of
stars \citep{bonnell01,bate03,wang09}.
Therefore, it is important to investigate the formation and evolution 
of dense cores in the cluster environment.

Recent theoretical studies on core formation in the cluster environment
have emphasized the role of supersonic turbulence
\citep{klessen98,tilley04,vazquez05,dib07}.
These studies have demonstrated that dense cores are formed in 
regions compressed by converging turbulent flows. 
However, the origin of this supersonic turbulence and its role
in core formation are still a matter of debate.
Two main scenarios have been proposed.
In the first scenario, formation of dense cores
(and thus star formation) is considered to be completed 
only in one turbulent crossing time, i.e., one dynamical time,
so that the cluster forming regions are destroyed within one dynamical time
\citep{elmegreen00,hartmann01,heitsch08}.
In this case, the mass spectrum of dense cores is shaped
by the cascade of primordial interstellar turbulence that has 
a power law form \citep{padoan02}.
On the other hand, in the second scenario, 
star formation is considered to continue at least over several dynamical times.
In this case, additional turbulent motions must be replenished
to keep supersonic turbulence in molecular clouds
because supersonic turbulence dissipates quickly in one dynamical time 
\citep{stone98,maclow99}.
In such circumstances, dense cores are continuously created 
(and destroyed) in molecular gas disturbed strongly 
by the additional turbulent motions.
In the cluster environment, stellar feedback such as stellar winds, 
HII regions, and protostellar outflows has been discussed as the dominant 
source of additional supersonic turbulence
\citep{norman80,matzner07,li06}.
Other sources of turbulence such as supernova blastwaves, external cloud 
shearing, or converging flows of cloud formation may also play a role
\citep[e.g.,][]{mckee07}.
As for the protostellar outflow-driven turbulence, \citet{swift08} 
identified the characteristic energy injection scale 
at $\approx$ 0.05 pc for the L1551 star forming region where
a group of stars are forming, indicating that the protostellar outflows
 are the primary source of regenerating turbulence in 
this region.
Recently, \citet{offner08} performed three-dimensional (3D)
hydrodynamic simulations of turbulent molecular clouds and 
suggested that the physical properties 
of dense cores may be affected directly by the presence or absence of 
turbulent feedback, although further justification may be needed.
Therefore, to shed light on the issue of core formation in the cluster
environment, it is essential to investigate the kinematics of
dense cores.
In this paper, we thus analyze the physical properties of 
dense cores in the nearest cluster forming region, 
the $\rho$ Ophiuchi molecular cloud.

The $\rho$ Ophiuchi molecular cloud is the nearest cluster forming 
regions at a distance of $\sim 125 $ pc 
\citep[e.g.,][]{lombardi08,loinard08,wilking08}.
The dense part of the $\rho$ Ophiuchi molecular cloud
(hereafter the $\rho$ Ophiuchi main cloud) is known to 
harbor a rich cluster of young stellar objects (YSOs)
in different evolutionary stages \citep{wilking08,enoch08,jorgensen08}. 
Recent millimeter and submillimeter dust continuum observations 
of the $\rho$ Ophiuchi main cloud have revealed that a large number of 
dense cores are concentrated in the central dense part of the cloud
and their mass spectra are similar in shape to the stellar IMF
\citep{motte98,johnstone00}.
By comparing with isothermal, pressure-confined, self-gravitating
Bonnor-Ebert spheres, \citet{johnstone00} argued that 
the dense cores detected by the dust continuum emission are likely to be 
self-gravitating and expected to form stars
\citep[see also][]{johnstone04}.
However, to accurately assess the dynamical states of the dense cores, 
it is necessary to perform molecular line observations using high density 
tracers such as H$^{13}$CO$^+$ ($J=1-0$) and N$_2$H$^+$
($J=1-0$), both of which have the critical densities for excitation
as high as $10^5$ cm$^{-3}$.
Recently, the Nobeyama observatory has released the archive data of 
the H$^{13}$CO$^+$ ($J=1-0$) line emission toward the $\rho$ Ophiuchi
main cloud taken with the 45 m radio telescope.
At a distance of 125 pc, the spatial resolution of the data of
21'', corresponding to $\sim 0.01$ pc, is 
comparable to those of the dust continuum observations, 
e.g., the SCUBA at the JCMT with a beam size 14''at 850 $\mu$m. 
Thus, the Nobeyama archive data is suitable for 
analyzing the dynamics of small scale structures 
comparable to the ones detected by the dust continuum observations.
Our analysis is complementary to that of \citet{andre07} who
observed several subclumps (Oph A, B1, B2, C, E, and F)
with the IRAM 30 m radio telescope using N$_2$H$^+$ ($J=1-0$) line 
to measure the line widths of the dust cores detected by
\citet{motte98}.
The H$^{13}$CO$^+$ data of the Nobeyama 45 m telescope
was taken in a larger area covering the whole dense region of 
the $\rho$ Ophiuchi main cloud, 
including all the subclumps observed by \citet{andre07}.
Since the dense gas detected by molecular line emissions
such as H$^{13}$CO$^+$ ($J=1-0$) and N$_2$H$^+$ ($J=1-0$)
may not be always well associated with the dust cores
and therefore the line widths of the dust cores estimated by 
\citet{andre07} may not be accurate, we identified the dense cores directly 
from the molecular line data cube
and estimated their physical quantities such as the core radius, 
mass, and line width.

The rest of the paper is organized as follows.
First, the description of the archive data is presented in \S \ref{sec:data}.
In \S \ref{sec:global} we show the spatial distribution 
of the dense gas detected by the H$^{13}$CO$^+$ emission,
which indicates very clumpy structures.
Then, in \S \ref{sec:core} we identify the clumpy structures as 
dense cores applying a clump-finding algorithm, 
the so-called clumpfind \citep{williams94}, and 
derive the physical properties of the identified cores.  
To accurately evaluate the core masses,
the fractional abundances of H$^{13}$CO$^+$ relative to H$_2$
are derived directly by comparing the H$^{13}$CO$^+$ column densities 
with the H$_2$ column densities  obtained from the 850 $\mu$m 
dust continuum emission.
In \S \ref{sec:discussion} we discuss the mass spectrum of 
the identified cores and their dynamical states. 
We also compare the physical properties of the identified cores 
with those in Orion A, taking into account the effect of 
different distances (or different spatial resolutions).
Finally, we summarize our main conclusion in \S \ref{sec:summary}.

%\section{Observations}
\section{Data}
\label{sec:data}

We used the archive data of the $\rho$ Ophiuchi main cloud 
taken in H$^{13}$CO$^+$ ($J=1-0$) molecular line (86.75433 GHz)
with the Nobeyama 45 m radio telescope.
The fits data is available from the web page of the Nobeyama Radio 
Observatory at the National Astronomical Observatory of Japan 
(\verb"http://www.nro.nao.ac.jp/"). 
The observations were carried out in the period from 2002 March to 2003 May.
All spectra were obtained in the position-switching mode.
The data has $118\times 98\times 60$ grid points in the 
$\alpha$-$\delta$-$v_{\rm LSR}$ space, covering the whole dense region
of the $\rho$ Ophiuchi main cloud with size of about $1.5$ pc $\times 1.3$ pc.
The grid size of the data was 21'', corresponding 
to 0.013 pc at a distance of 125 pc.
The main beam efficiency of the telescope was 0.51.
We note that the FWHM beam size of the telescope was 18''
at 87 GHz, and therefore the data were almost full-beam sampled,
considering the typical pointing error of a few arcsecond
(see the web page of the Nobeyama Radio Observatory).
The velocity resolution is 0.13 km s$^{-1}$
and the average Root Mean Square (rms) noise  determined 
from signal-free channels is $\sigma=$ 0.11 K in $T_A^*$.

%\section{Results}
\section{Distribution of Dense Molecular Gas 
in the $\rho$ Ophiuchi molecular cloud}
\label{sec:global}

\subsection{Overall Distribution of the H$^{13}$CO$^+$ emission}

\begin{figure}
\epsscale{0.8}
\plotone{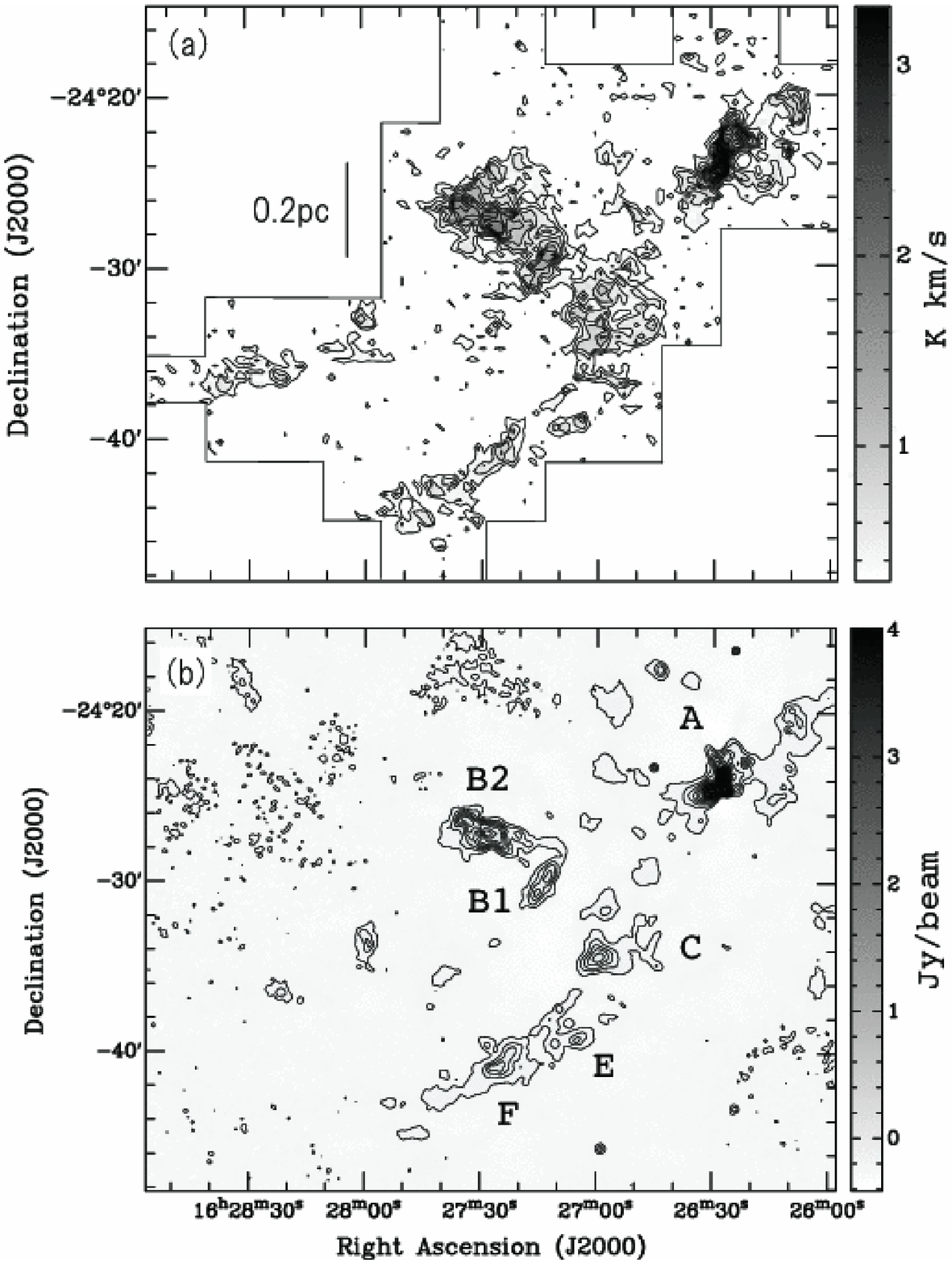}
\caption{(a): H$^{13}$CO$^+$ ($J=1-0$) total integrated intensity map 
in the velocity range of $v_{\rm LSR}=1-6$ km s$^{-1}$ 
toward the $\rho$ Ophiuchi main cloud.
The grey scale indicates the integrated intensity in units of K km s$^{-1}$.
The contours start from 0.3 K km s$^{-1}$, corresponding to the 3 $\sigma$
noise level,  at intervals of 0.2 K km s$^{-1}$. 
(b): 850$\mu$m image of the same area as in panel (a),
 obtained with the SCUBA at the JCMT.
The grey scale indicates the intensity in linear scale
from -0.4 to 4 Jy beam$^{-1}$. 
The contours start from 0.2 Jy beam$^{-1}$ at intervals of 0.2 Jy beam$^{-1}$.
The dense subclumps are designated by A, B1, B2, C, E, and F.
}  
\label{fig:global}
\end{figure}

Here, we present the overall distribution of dense
gas observed by the H$^{13}$CO$^+$ ($J=1-0$) emission.
Figure \ref{fig:global}a shows a total integrated intensity map 
of the H$^{13}$CO$^+$ emission toward the $\rho$ Ophiuchi main cloud.
For comparison, the 850 $\mu$m image obtained 
with the SCUBA at the JCMT is shown in Figure \ref{fig:global}b. 
While the distribution of the H$^{13}$CO$^+$ emission 
appears to cover a larger area than that of the 850 $\mu$m emission,
the overall spatial distribution of the H$^{13}$CO$^+$ emission
is similar to that of the 850 $\mu$m emission.
We note that it is very difficult to observe extended structures with 
bolometers mounted on ground-based telescope due to the atmospheric
emission.  The extended structures larger than $\sim$ 1.5'
($\approx 0.06$ pc) are therefore suppressed in removing the atmospheric 
emission during the data reduction process,
 making the 850 $\mu$m map mostly devoid of extended emission
\citep{johnstone00b}.
Our H$^{13}$CO$^+$ map shows that the well-known dense
subclumps (Oph A, B1, B2, C, E, and F. Note that 
the Oph D region is out of our area.), 
previously identified by DCO$^+$ and other observations,
are very clumpy as recent submillimeter observations 
have revealed \citep{motte98,johnstone00}.
In the next section, we identify the clumpy structures as 
dense cores applying a clump-finding algorithm, 
 clumpfind \citep{williams94}, 
and derive physical properties of the identified cores.

\subsection{Individual Regions}

\begin{figure}
\epsscale{1.0}
\plottwo{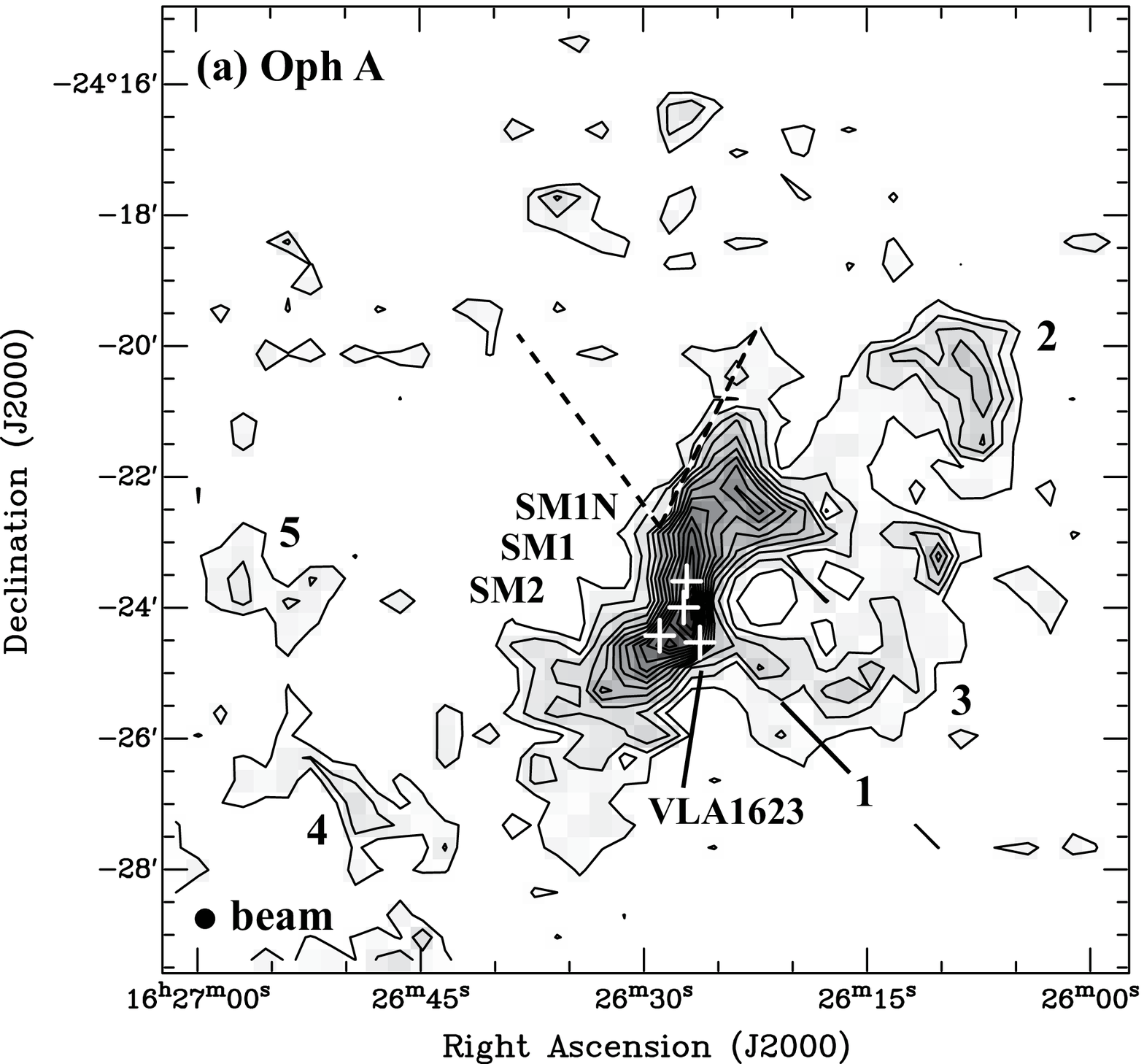}{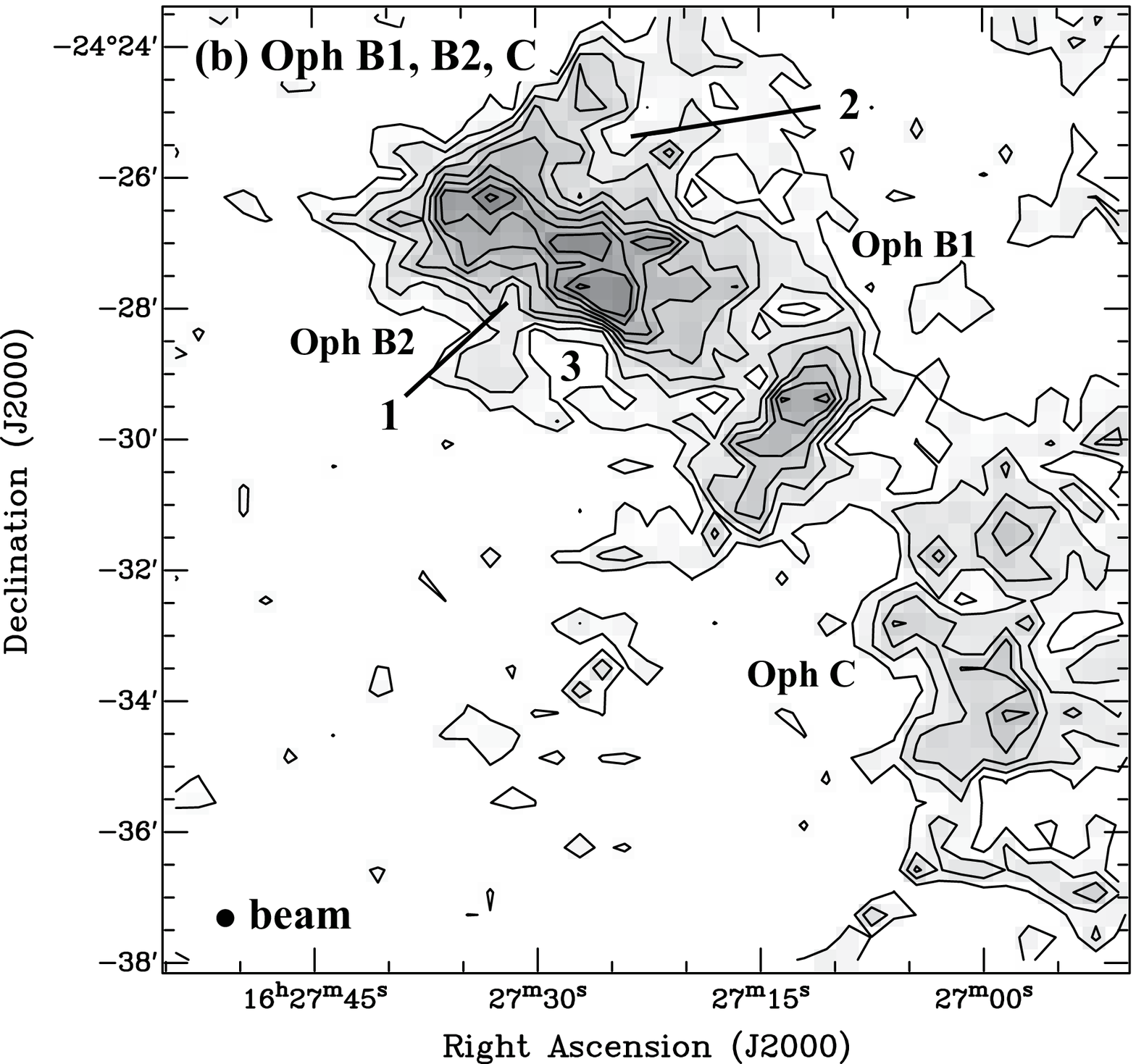}
\caption{H$^{13}$CO$^+$ ($J=1-0$) total integrated intensity maps 
taken with the Nobeyama 45 m telescope toward (a) the Oph A region
 and (b) the Oph B1, B2, and C regions.
The grey scale and contours are the same as those of Figure \ref{fig:global}a.
In  panel (a), the two dahed lines
% that intersect at (16:26:29,  -24:22:45) 
indicate the position of the filaments seen in
the 850$\mu$m  map \citep{wilson99}.
The positions of the submillimeter sources, SM1, SM1N, and SM2, and the
 prototypical Class 0 object, VLA 1623, are indicated by crosses.
In each panel the arcs and holes
discussed in the text are indicated by numbers.
The beam size of the Nobeyama 45 m  telescope is shown 
at the lower left corner of each panel.
}  
\label{fig:ophabd}
\end{figure}

\begin{figure}
\epsscale{0.5}
\plotone{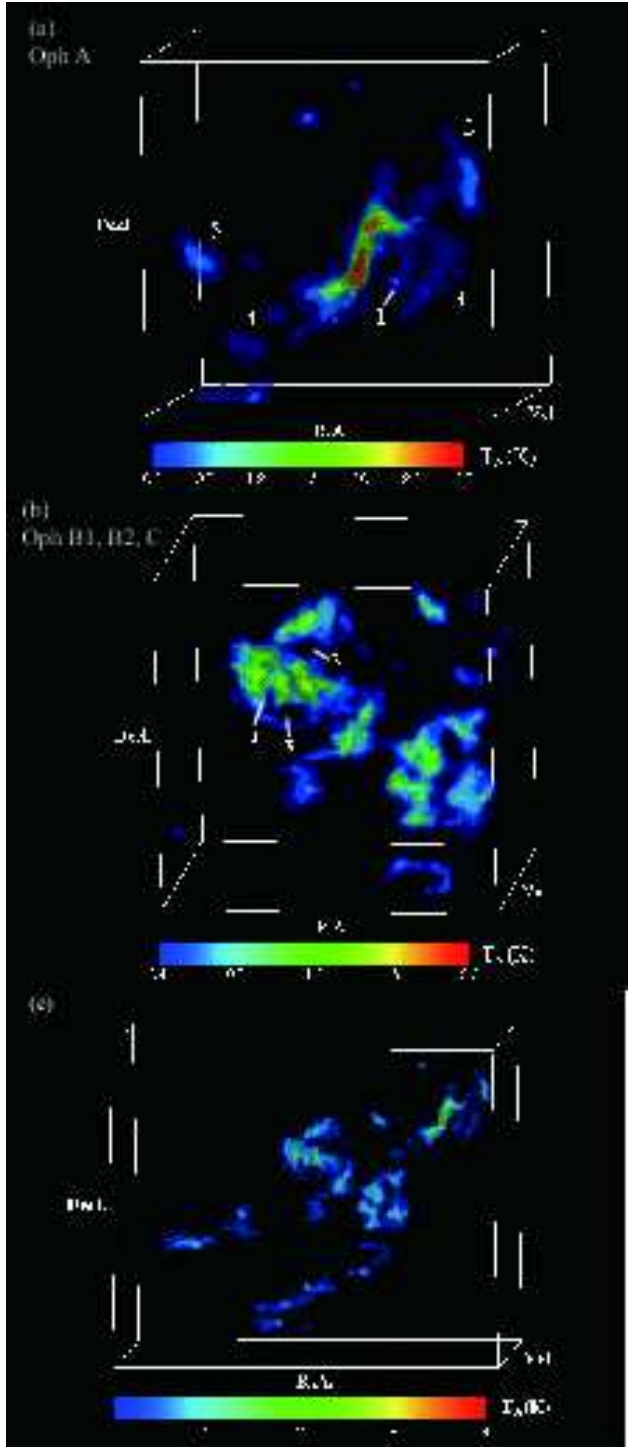}
\caption{Three-dimensional representation of the antenna temperature ($T_A^*$)
in the $\alpha$-$\delta$-$v_{\rm LSR}$ space 
toward (a) the Oph A region, (b) the Oph B1, B2, and C regions,
and (c) the whole observed area.
For panels (a) and (b), the areas projected on the plane of the sky 
are the same as those of Figures \ref{fig:ophabd}a and
 \ref{fig:ophabd}b, respectively.
The color shows the iso-temperature surfaces.
The color bar indicates the antenna temperature in units of K.
In each panel the numbers have  the same meaning as in Figure \ref{fig:ophabd}.
(Mpeg animations of Figs. \ref{fig:ophabd}a, \ref{fig:ophabd}b, and 
\ref{fig:ophabd}c are available in the online journal.)
}  
\label{fig:ophabd3d}
\end{figure}

Blow-up total integrated intensity maps toward the Oph A and Oph B1, B2 and C
regions are shown in Figures \ref{fig:ophabd}a and \ref{fig:ophabd}b, 
respectively. Iso antenna temperature surfaces 
of the same areas in the 3D data space ($\alpha$-$\delta$-$v_{\rm LSR}$) 
are also shown in Figure \ref{fig:ophabd3d}.
Our intensity maps can be compared 
with Figure 2 of \citet{andre07}, 
the N$_2$H$^+$ ($J=1-0$) intensity maps obtained with the IRAM 30 m telescope.
The beam sizes for both the telescopes are comparable: 
the HPBW of the IRAM 30 m, 26.4'', is about 1.5 times 
as large as that of the Nobeyama 45 m telescope, 18''.
Both the molecules are also expected to trace dense molecular gas in 
nearly the same density range 
because the critical densities for excitation of H$^{13}$CO$^+$ ($J=1-0$) 
and N$_2$H$^+$ ($J=1-0$) lines are comparable
 ($n_{\rm cr, H^{13}CO^+} = 8\times 10^4$ cm$^{-3}$ and
$n_{\rm cr, N_2H^+} = 2\times 10^5$ cm$^{-3}$).
In fact, the distribution of the H$^{13}$CO$^+$ emission is remarkably 
similar to that of the N$_2$H$^+$ emission.

The most intense line emission of H$^{13}$CO$^+$ ($J=1-0$) comes from 
the Oph A region.
The maximum antenna temperature reaches $T^*_A \simeq 4$ K
at the position that coincides with that of a starless object, SM1, 
identified by \citet{andre93} within our spatial resolution.
A prominent feature of this region is a dense ridge
running from north to south, which is also seen 
in the 850 $\mu$m map.
The starless objects SM1, SM1N, and SM2  identified by \citet{andre93} 
are located just at the densest part of the ridge.
Although relatively strong H$^{13}$CO$^+$ emission is detected toward 
VLA1623, the prototypical Class 0 object, it is difficult to distinguish
 the H$^{13}$CO$^+$ emission associated with VLA 1623 from the ambient 
component because of our resolution as large as 21''.

In the 850 $\mu$m map, \citet{wilson99} found 
two filaments in the Oph A region [see Figure 1 of \citet{wilson99}].
Similar structures can be recognized in our H$^{13}$CO$^+$ map.
In the 850 $\mu$m map, the two filaments
(the dashed lines in Figure \ref{fig:ophabd}a)
intersect at the position (RA, Dec) = 
$(16^{\rm h}26^{\rm m}29^{\rm s}, -24^\circ22'45''; {\rm J2000})$.
In the H$^{13}$CO$^+$  map, the north-east filament
 is not seen, whereas the north-west filament,
 a part of the Oph A ridge, has strong H$^{13}$CO$^+$ emission.
Such a feature is consistent with that of the N$_2$H$^+$ 
map by \citet{james04}, 
indicating that the north-east filament presumably consist of diffuse warm gas.
The filament is reminiscent of the Orion bar
where the dust continuum emission enhanced by heating from PDRs and 
young stars is prominent but the H$^{13}$CO$^+$ line emission is not seen 
\citep{ikeda07}. In fact, the location of the north-east filament 
agrees well with the edge of the PDR seen in ISO images \citep{abergel96}.

In the western side of the Oph A ridge, \citet{wilson99}
found another interesting structure: 
two arcs that are labeled with No. 1 and 2 in 
Figures \ref{fig:ophabd}a and \ref{fig:ophabd3d}a. 
In addition to these two, we identified 3 new arcs 
in the Oph A region (labeled with No. 3 through 5). 
Although the No. 1 arc is not clearly recognized as an arc 
in the total intensity map of Figure \ref{fig:ophabd}a 
owing to the presence of the extended diffuse component,
the No. 1 arc as well as the No. 2 and 3 can be clearly 
seen in the iso-temperature surfaces in Figure \ref{fig:ophabd3d}a.
These arcs may be created by the stellar wind from the young B3
star S1.
Another possibility is the effect of three outflows
detected in this region whose outflow axes are almost 
parallel in the $\alpha$-$\delta$ space 
and inclination angles are around 60 $-$ 80 degrees \citep{kamazaki03}.
One of the three outflows is the most spectacular outflow 
from the Class 0 object VLA 1623. 
Both sides of the No. 1 arc extended from VLA 1623, 
where no H$^{13}$CO$^+$ emission is detected, appear to be parts of a cavity 
created by the VLA 1623 outflow. 
On the other hand, in the eastern side of the Oph A ridge, 
we can recognize two new arcs.
These arcs (labeled with No. 4 and 5) appear to be related to a 
giant protostellar outflow that has been recently discovered by
Nakamura et al. (2009, in preparation) on the basis of the $^{12}$CO
($J=3-2$) observations.

A couple of features that are likely to be created by protostellar outflows
are also seen in the Oph B2 region, the densest region after the Oph A.
The H$^{13}$CO$^+$ map shows a very clumpy structure 
with $\sim$ 10 local peaks in the region.  
Many dust cores identified by \citet{motte98} 
are distributed mainly in the bright area
of the H$^{13}$CO$^+$ map, where
some cores identified by the clumpfind in the data cube 
(see \S \ref{sec:core}) are overlapped on 
the plane of the sky. 
It is difficult to distinguish such an overlapped core 
from the other in the H$^{13}$CO$^+$ total integrated intensity map 
or the dust continuum map that 
does not have velocity information.
In the southern part of the Oph B2 region (labeled with No. 1), 
a hole is seen in both the H$^{13}$CO$^+$ and 
850 $\mu$m maps.  The hole is particularly clear in the data
cube in Figure \ref{fig:ophabd3d}b.
Two other holes are also seen near the northern (No. 2) and 
southern (No. 3) edges of the B2 region.
%fn
These holes are likely to be created by the giant protostellar
outflow recently found by Nakamura et al. (2009).
The blue lobe of this outflow is extended beyond the B2 region,
reaching the No. 4 and 5 arcs. The apparent luminosity 
and the total mass of this outflow are largest in $\rho$ Oph.
The driving source is likely to be a class I YSO Elias 32 or 33.
No. 1 and 2 holes are likely to be created by the red and blue lobes,
respectively.
The interaction between the protostellar outflows and the dense gas
will be discussed elsewhere (Nakamura et al. 2009, in preparation). 
%f
%These holes could be created by protostellar
%outflows, although the driving sources are unknown.
%Some of YSOs detected by the Spitzer Space telescope 
%might be responsible for the formation of the holes
%(see Figure \ref{fig:stardistribution}).
Cavities that could be created by protostellar outflows 
are also found in NGC1333, a nearby cluster forming clump
in the Perseus molecular cloud \citep{knee00,sandell00,quillen05},
suggesting that density structures in 
cluster forming regions are strongly affected by protostellar outflows.
Nakamura et al. (2009) also found other examples
of the dynamical interaction between the dense gas and the outflows
in $\rho$ Oph, which will be shown elsewhere.

\section{Physical Properties of Dense Cores}
\label{sec:core}

We identified 68 H$^{13}$CO$^+$ cores in the $\rho$ Ophiuchi main cloud, 
using the clumpfind method described in \S \ref{subsec:identification}.
Here, we describe physical properties of the identified cores.
The definitions of the physical quantities of the cores are presented in
\S \ref{subsec:derivation}. 
The fractional abundance of H$^{13}$CO$^+$ relative to H$_2$,
which is important to the determination of core mass, is 
estimated in \S \ref{subsec:abundance}.
In \S \ref{subsec:distribution}, we describe the statistical properties 
of the physical quantities of the identified cores, and 
compare their positions with those of YSOs identified 
by the Spitzer Space Telescope mid-infrared data
\citep{jorgensen08}.
We also compare our cores with 850 $\mu$m cores
listed in Table 7 of \citet{jorgensen08}. 
Finally, we show in \S \ref{subsec:correlation} the 
correlation among physical quantities such as the 
line width-radius and virial ratio-mass relations.

\subsection{Core Identification}
\label{subsec:identification}

To identify dense cores from the H$^{13}$CO$^{+}$ 
 data cube, we applied the clumpfind method 
described in \S 3.2 of \citet{ikeda07}. 
%fn
In brief, we adopted the threshold ($T_{A, \rm th}$) and stepsize 
($\Delta T_A$) of the 2 $\sigma$ noise level for core identification, 
following  \citet{williams94} who recommended using this set of the parameters 
based on the results of their test simulations.
To identify the cores with reasonable accuracy, 
we imposed the following 3 criteria: 
(1) a core must contain two or more continuous velocity channels, 
each of which has at least 3 pixels whose intensities are 
above the 3$\sigma$ noise level, and (2) the pixels
must be connected to one another in both the space and velocity
domains. (3) Furthermore, we rejected core candidates located at the edge 
of the observed region.
The total intensity of the minimum mass core identified 
by this method has about 60 $\sigma$ levels.
Using our definition of core mass (eq.[\ref{eq:mlte}]), 
a minimum core mass is estimated 
to be 0.28 M$_\odot$ and 0.38 M$_\odot$ for $T_{\rm ex}=12$ K
and 18 K, respectively.  
The total mass of the identified 68 cores is estimated to be 
228 $M_\odot$, 60 \% of the total mass detected above the $2\sigma$ 
level.

In general, the physical properties of the identified cores depend on the core 
identification scheme and its parameters. In our case, we expect that
our clumpfind can detect the real 3D density structures in the cloud
from the H$^{13}$CO$^+$ ($J=1-0$) spectral line
position-position-velocity data cube because of the following reasons.
According to \citet{williams94}, the physical properties 
of the cores identified by clumpfind are affected siginificantly by 
the adopted threshold and stepsize unless they are carefully chosen.
They recommend using the 2 $\sigma$ noise level for both the 
threshold and stepsize, for which the clumpfind could detect and 
accurately measure the individual core properties in their test
simulations. We adopt this set of the parameters in our clumpfind.
Recently, several authors revisted this problem \citep{pineda09,ikeda09b}.
\citet{pineda09} reexamined the effect of 
varying the values of the threshold ($T_{A, \rm th}=$3, 5, and 7
$\sigma$) 
and stepsize ($3 \sigma \le \Delta T_A \le 20 \sigma$) 
for the core identification, 
and confirmed that the physical 
quantities of the identified cores depend on these parameters 
especially when the distribution of the emission is not 
localized in the data cube and the larger values of the 
parameters are used [see also \citet{schneider04} for 
the case of the low-spatial resolution ($\sim $1 pc), 
optically thick $^{12}$CO ($J=1-0$) data 
with $T_{A, \rm th}=\Delta T_A=$ 3 $\sigma$].
\citet{ikeda09b} explored more reasonable parameter ranges of 
$T_{A, \rm th}=2-3 \sigma$ and $2 \sigma \le \Delta T_A \le 3 \sigma$ 
suitable for their core study in the OMC-1 region, and 
demonstrated that the power-law index of a core mass spectrum
does not depend on the threshold and stepsize.
\citet{williams94} also discussed the ``overlap'' effect for which 
multiple cores that almost overlap along the line-of-sight
are not separately identified by the clumpfind when they have
nearly the same line-of-sight velocities. 
This overlap effect is likely to be significant when 
the gas is highly turbulent and  the emission tends
to be distributed in the large volume of the 3D data cube
\citep{ballesteros02,smith08}.
For example, \citet{ostriker01} performed 3D MHD turblent simulations
 and claimed that  the emission lines are often singly peaked
even in the presence of multiple condensations along the line of sight. 
We note that their model clouds are highly turbulent, gravitationally 
unbound and the condensations have extremely large 
virial ratios, ranging from 10 to $10^3$. 
This suggests that the overlap effect may be crucial for less dense 
gravitationally unbound molecular gas components 
observed by $^{12}$CO and $^{13}$CO
\citep[see][for the case of the Polaris molecular cloud,
a nearby diffuse, gravitationally unbound molecular cloud.]{ossenkopf02}.
In contrast, for our H$^{13}$CO$^+$ data, the overlap effect 
is expected to be minor because the H$^{13}$CO$^+$ emission 
can trace only dense gas with densities greater than $\sim 10^5$ cm$^{-3}$
and appears to be reasonably localized in the 3D data cube,
indicating that 
%.
%This is due in part to the fact that 
the dense gas is likely to have a small volume filling factor 
in the cloud and moderately quiescent 
(the Mach number $\sim 2$). 
Note that the low density gas has the Mach number of 
about 5 $-$ 6 \citep{loren89}.
In fact, our cores have virial ratios of about 2, 
as shown in \S \ref{subsub:yso}. Thus, 
we expect that the clumpfind can identify the 3D density structures
reasonably well from our H$^{13}$CO$^+$ data.
%f

A caveat of our analysis is that the cores 
identified by the above procedure (and any other 
core identification schemes) are 
resolution dependent in the sense that the substructures 
smaller than the telescope beam size cannot be resolved.
It remains unclear whether an identified core is a unit that is 
separated dynamically from the background media.
However, we expect that the structures identified from 
high density tracers such as H$^{13}$CO$^+$ ($J=1-0$), N$_2$H$^+$ ($J=1-0$),
and dust continuum data correspond to the real dense 
structures in the cloud as long as the spatial resolution is 
less than $\sim$ 0.01 pc,
because recent numerical simulations of pc-scale cluster forming clumps 
demonstrated that the dense structures like filaments often created in 
the clump have sizes of order 0.1 pc or less along the minor axes 
and densities of order $10^5$ cm$^{-3}$ \citep{li06,nakamura07}.
Therefore, in the following, we discuss the statistical 
properties of the cores identified by the above procedure, assuming that
the identified cores can be regarded as units that are reasonably separated 
dynamically from the background media.
Further investigation of this problem should be done when 
higher spatial resolution observations are available.
%f

\subsection{Derivation of Core Properties}
\label{subsec:derivation}

We derive physical properties of the cores using 
the definitions described in \S 3.3 of \citet{ikeda07}.
In brief, the position and local standard of rest (LSR) velocity
of each core are determined at the pixel with the largest
antenna temperature, $T_{A,\rm peak}^*$, within the core.
The core radius, $R_{\rm core}$, is determined by taking the projected 
area enclosed by the 2 $\sigma$ level contour 
and computing the radius of the circle required to reproduce the area, 
taking into account the correction for the telescope beam (18'').
The aspect ratio is computed as the ratio of the major to minor axis
lengths that are determined by the two-dimensional Gaussian fitting 
to the total integrated intensity distribution of the core.
The FWHM line width, $dv_{\rm core}$, is corrected for the 
velocity resolution of the spectrometers ($=0.13$ km s$^{-1}$)
and for line broadening due to hyperfine splitting of 
H$^{13}$CO$^+$. 
The LTE mass is estimated as 
\begin{eqnarray}
M_{\rm LTE}&=&1.88 \times 10^{-2} \left({X_{\rm H^{13}CO^+} \over
	 1.72\times 10^{-11}}\right)^{-1}
T_{\rm ex} \exp\left({4.16/T_{\rm ex}}\right) \nonumber \\
&&\times \left({D\over 125 {\rm pc}}\right)^2
\left({\Delta \theta \over 21\arcsec}\right)^2 
\left({\eta \over 0.51}\right)^{-1}
\left({\sum_i T_{A,i}^* \Delta v_i \over {\rm K \ km \ s^{-1}}}\right)
M_\odot , 
\label{eq:mlte}
\end{eqnarray}
by assuming that the H$^{13}$CO$^+$ emission is optically thin.
Here, $X_{\rm H^{13}CO^+}$ is the fractional abundance of H$^{13}$CO$^+$
relative to H$_2$, $T_{\rm ex}$ is the excitation temperature, $D$
is the distance to the $\rho$ Ophiuchi main cloud, 
$\sum_i T_{A,i}^* \Delta v_i$
is the total integrated intensity of the core and $\Delta v_i = 0.13$
km s$^{-1}$. 
We adopt the fractional abundance 
$X_{\rm H^{13}CO^+}=1.72\times 10^{-11}$ (see \S \ref{subsec:abundance}), 
and the distance $D=125$ pc 
on the basis of the values recently updated by 
\citet{lombardi08} and \citet{loinard08} \citep[see also][]{wilking08}.
The excitation temperature is assumed to be $T_{\rm ex}=12$ K \citep{motte98}
except for the area shown in Figure \ref{fig:ophabd}a,
i.e., the Oph A region, where $T_{\rm ex}=18$ K
following the N$_2$H$^+$ observations by \citet{james04}.

The virial mass $M_{\rm vir}$ is calculated as
\begin{eqnarray}
M_{\rm vir}&=&\frac{5a^{-1}R_{\rm core}dv_{\rm tot}^2}{8\ln 2G}
 \nonumber \\
&=& 209 a^{-1}\left({R_{\rm core}\over {\rm pc}}\right)
\left({dv_{\rm tot}\over {\rm km} \ {\rm s}^{-1}}\right)^2 M_\odot ,
\label{eq:virial mass}
\end{eqnarray}
and 
\begin{equation}
dv_{\rm tot} = \left[dv_{\rm core}^2 + 8\ln 2 \, k_BT
\left({1 \over \mu m_H}-{1\over m_{\rm obs}}\right)\right]^{1/2}
\end{equation}
where $k_B$ is the Boltzmann constant, 
$T$ is the kinetic temperature of the molecular gas,
$\mu$ is the mean molecular weight of 2.33, 
$m_H$ is the mass of a hydrogen atom, 
$m_{\rm obs}$ is the mass of a H$^{13}$CO$^+$ molecule, and 
$a$ is a dimensionless parameter of order unity which measures the
effects of a nonuniform or nonspherical mass distribution \citep{bertoldi92}. 
For a uniform sphere and a centrally-condensed sphere with 
$\rho \propto r^{-2}$, $a=1$ and 5/3, respectively.
For our cores, the effects of the nonspherical mass distribution appear to
be small because the aspect ratios are not so far from unity
(see also Figure 2 of Bertoldi \& McKee 1992).
The virial ratio, the ratio between the virial mass to the LTE mass, 
is defined as $\alpha_{\rm vir}\equiv M_{\rm vir}/M_{\rm LTE}$,
which is  equal to twice the ratio of the internal kinetic energy to 
the gravitational energy.

The mean number density of the core, $\bar{n}$, is calculated
as the LTE mass divided by the volume of a sphere with radius, 
$R_{\rm core}$,
\begin{equation}
\bar{n}={3M_{\rm LTE} \over 4\pi \mu m_H R_{\rm core}^3}  \ . 
\end{equation}

\subsection{Fractional Abundance of H$^{13}$CO$^+$}
\label{subsec:abundance}

\begin{figure}
\epsscale{1.0}
\plotone{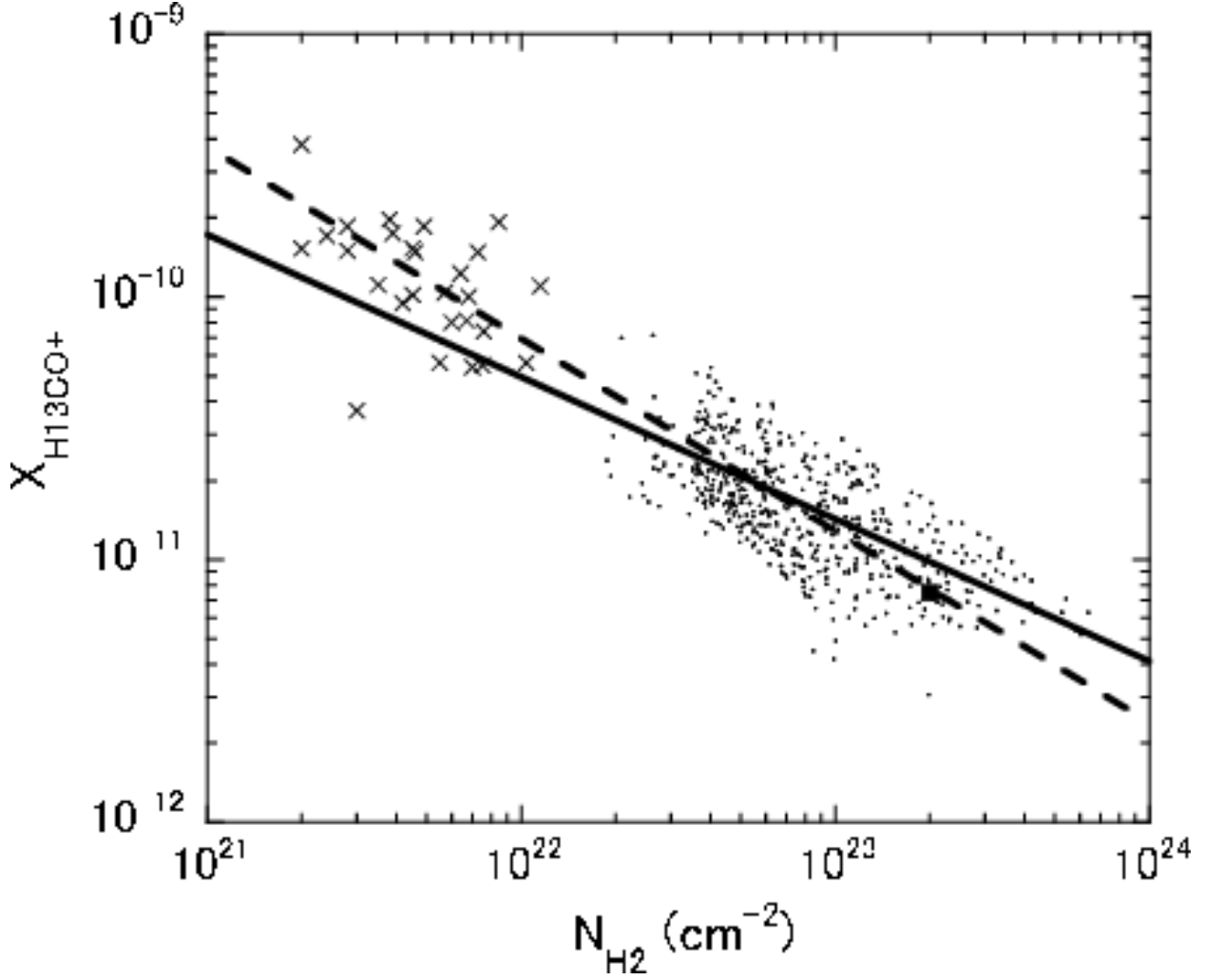}
\caption{Fractional abundance of H$^{13}$CO$^+$ 
relative to H$_2$ against the H$_2$ column density
in the $\rho$ Ophiuchi main cloud. 
The solid line shows the best-fit power-law for the $\rho$  Oph data.
For comparison, the fractional abundances of H$^{13}$CO$^+$
derived by \citet{butner95} and \citet{dishoeck95} are plotted by crosses and 
a filled square, respectively.
The dashed line shows the best-fit power-law for both
their data and ours.}
\label{fig:abundance}
\end{figure}

To evaluate the fractional abundance of ${\rm H^{13}CO^+}$,
we smoothed the 850 $\mu$m map with the angular resolution 
of 21'' to match the H$^{13}$CO$^+$ total integrated intensity map
and computed the column densities of H$^{13}$CO$^+$
($N_{\rm H^{13}CO^+}$) and H$_2$ ($N_{\rm H_2}$) at each pixel in the maps.
Here, we computed $N_{\rm H_2}$ from the relation 
\begin{equation}
N_{\rm H_2}=S^{\rm beam}_{850}/[\Omega_{\rm beam}\mu m_H
\kappa_{850}B_{850} (T)] , 
\end{equation}
where
$S^{\rm beam}_{\rm 850}$ is the 850 $\mu$m flux density per beam,
$\Omega_{\rm beam}$ is the main beam solid angle, 
$B_{850}(T)$ is the Plank function for a dust temperature $T$, 
 and 
$\kappa_{850}$ (= 0.01 cm$^2$g$^{-1}$) is the dust opacity at 850 $\mu$m
 \citep{henning95,ossenkopf94,johnstone00}.
The dust temperature is assumed to be equal to the gas 
temperature.
In Figure \ref{fig:abundance} the fractional abundances 
$X_{\rm H^{13}CO^+}$ computed at all the pixels above the 3 $\sigma$ 
levels for both the H$^{13}$CO$^+$ and 850 $\mu$m maps
are plotted against $N_{\rm H_2}$.
We note that the H$_2$ column density derived 
from the 850 $\mu$m map is very sensitive to the 
temperature and the dust opacity assumed, 
and therefore have uncertainty of at least a factor of a few.
The fractional abundance of H$^{13}$CO$^+$ tends to decrease
with increasing H$_2$ column density. 
The best-fit power-law is given by 
\begin{equation}
\log X_{\rm H^{13}CO^+} = (-10.87\pm 0.01)
+ (-0.514\pm 0.024) 
\log(N_{\rm H_2}/10^{23} {\rm cm^{-2}})
\label{eq:abundance}
\end{equation}
with a correlation coefficient ${\cal R}=0.68$, indicating 
that $X_{\rm H^{13}CO^+}$ is nearly proportional to $N_{\rm H_2}^{-1/2}$.
This tendency is in good agreement with a theoretical 
consideration of the ionization fraction in  molecular clouds
if the column density is proportional to the local volume density
for each molecule.
When the ionization rate by cosmic rays 
($\propto n_n$, where $n_n$ is the density of neutral gas)
balances with the recombination rate ($\propto n_i n_e$, where
$n_i$ and $n_e$ are the densities of positive ion and electron,
respectively, and $n_i \approx n_e$), the ionization fraction is inversely 
proportional to the square root of the neutral gas density
\citep{elmegreen79}. 
In typical regions in molecular clouds, the most abundant
positive molecular ion is expected to be HCO$^+$. 
Therefore, the fractional abundance of H$^{13}$CO$^+$
 as well as that of HCO$^+$ is expected to be  
inversely proportional to the square root of the neutral gas density
if the fractional abundance of H$^{13}$CO$^+$ relative to 
HCO$^+$ is almost constant.
More detailed studies indicate that the relation
$X_{\rm H^{13}CO^+}\propto n_n^{-1/2}$ is a reasonable approximation 
of the fractional abundance in the density range 
of $n\lesssim 10^7$ cm$^{-2}$, 
although metal ions such as Mg$^+$ may be the most abundant positive ions
\citep{nishi91, nakano02}.
We note that as mentioned in \S \ref{sec:global}, in the 850 $\mu$m data 
the structures larger than 1.5' ($\approx 0.06$ pc) are removed during
the data reduction process. To evaluate how the effect of this 
artificial filtering affects the estimate of the 
H$^{13}$CO$^+$ abundance, we repeated the same analysis
by filtering out all the structures larger than 1.5' in 
the H$^{13}$CO$^+$ map and confirmed that the effect of 
this artificial filtering is negligible.
This is because the dense gas detected by the 850 $\mu$m
and H$^{13}$CO$^+$ is spatially well-localized.

In nearby low-mass star forming regions, \citet{butner95} found 
that $X_{\rm H^{13}CO^+}$ ranges from $3\times 10^{-11}$ 
to $4\times 10^{-10}$ for dense cores 
with $2\times 10^{21}$ cm$^{-2}$ $\lesssim N_{\rm H_2}
\lesssim 1\times 10^{22}$ cm$^{-2}$, 
where they estimated the H$_2$ column densities 
from C$^{18}$O ($J=1-0$) observations. 
For comparison, their values are indicated by crosses 
in Figure \ref{fig:abundance}. 
Although their $X_{\rm H^{13}CO^+}$ tends to decrease 
with increasing $N_{\rm H_2}$ in the same way as for our data, 
their $X_{\rm H^{13}CO^+}$ tend to be somewhat larger than the values 
predicted from eq. [\ref{eq:abundance}].  
If $X_{\rm H^{13}CO^+}$ derived by \citet{butner95} 
is fitted by a power-law with the same power index 
as in eq. [\ref{eq:abundance}], 
the coefficient becomes about twice 
that of eq. [\ref{eq:abundance}].
It is difficult, however, to judge whether this difference 
originates from the different environments 
because both the data have uncertainty
at least by a factor of a few.
If both the data are fitted by 
a power-law with a single index, then 
the best-fit function is given by 
\begin{equation}
\log X_{\rm H^{13}CO^+} = (-10.89\pm 0.01)
+ (-0.732\pm 0.014) 
\log(N_{\rm H_2}/10^{23} {\rm cm^{-2}})
\label{eq:abundance2}
\end{equation}
with ${\cal R}=0.87$. 
The mean fractional abundance $\left<X_{\rm H^{13}CO^+}\right>$
is estimated to $1.72 \times 10^{-11}$ 
with a standard deviation of $0.92\times 10^{-11}$. 
In the present paper, we assume the constant fractional abundance of 
$X_{\rm H^{13}CO^+} = 1.72 \times 10^{-11}$ for the entire area.
In future analysis, the spatial variation 
in fractional abundance will be taken into account.

\citet{dishoeck95} measured a smaller fractional abundance  
$X_{\rm H^{13}CO^+}\approx 0.75\times 10^{-11}$ 
toward IRAS 16293$-$2422, a young class 0 binary YSO, 
located in the eastern streamer region of the $\rho$ Ophiuchi molecular 
cloud complex \citep[see][]{andre00}. 
%fn
They obtained the H$_2$ column density of $2\times 10^{23}$ cm$^{-2}$
toward this YSO from the C$^{17}$O observations, somewhat larger than
 the mean H$_2$ column density of $\rho$ Oph evaluated 
from the 850 $\mu$m data.
Their value, plotted in Fig. \ref{fig:abundance} by a filled square, 
is almost on our best-fit power-law of eq. [\ref{eq:abundance2}].
%Although their value is within the distribution of our estimated 
%value, it is unlikely that the mean 
%$X_{\rm H^{13}CO^+}$ toward the $\rho$ Ophiuchi 
%main cloud is as small as $0.75\times 10^{-11}$.
%If $\left<X_{\rm H^{13}CO^+}\right>\approx 0.75\times 10^{-11}$, the total
%gas mass detected by the H$^{13}$CO$^+$ emission 
%($\approx 592 M_\odot$) becomes larger than  
%that detected by a lower density tracer C$^{18}$O of 494M$_\odot$
%(see \S \ref{sec:global}).
Thus, we believe that our adopted fractional abundance 
of $ 1.72\times 10^{-11}$
is a plausible representative value 
for the $\rho$ Ophiuchi main cloud.

We note that several high-resolution interferometric observations of 
starless cores in relatively quiescent, low temperature 
star forming regions like Taurus have revealed that 
several gas-phase molecules including CO, HCO$^+$, and their isotopes tend to 
deplete at the densities higher than $10^5$ cm$^{3}$.
One might interpret that the decrease in the H$^{13}$CO$^+$ abundance 
with increasing column density is due to the effect 
of the depletion. However, the effect of the depletion should  
be small because of the following reasons.
In cluster forming regions like $\rho$ Oph, relatively high 
turbulent motions, protostellar outflows and stellar radiation
tend to increase the temperature steadily and/or temporarily, preventing 
or slowing down the molecular depletion in starless cores. 
In addition, the high density part where the molecular depletion is 
significant is spatially localized and therefore the interferometric 
observations with high spatial resolution are needed to resolve the
effect of depletion in a core. However, our beam size
($\sim 20''$) is not enough to well resolve the structures inside 
each core \citep{aikawa95}.
Furthermore, even if the depletion of gas-phase CO and HCO$^+$ is 
significant in $\rho$ Oph, the fractional abundance of HCO$^+$
relative to CO does not change with increasing
column density because these carbon-bearing molecules tend to 
equally deplete \citep{aikawa95}. 
In contrast, the H$^{13}$CO$^+$ fractional abundances of
\citet{butner95} and \citet{dishoeck95}, for both of which 
the H$_2$ column densities are measured from the CO isotopes, 
do not show such a 
constant abundance, but decrease significantly with increasing column density,
indicating that the dependence of the H$^{13}$CO$^+$ abundance on the
column density comes from the different physics
from the molecular depletion.

\subsection{Distribution of Physical Quantities of Dense Cores and 
Comparison with Dust Cores}
\label{subsec:distribution}

\begin{figure}
\plotone{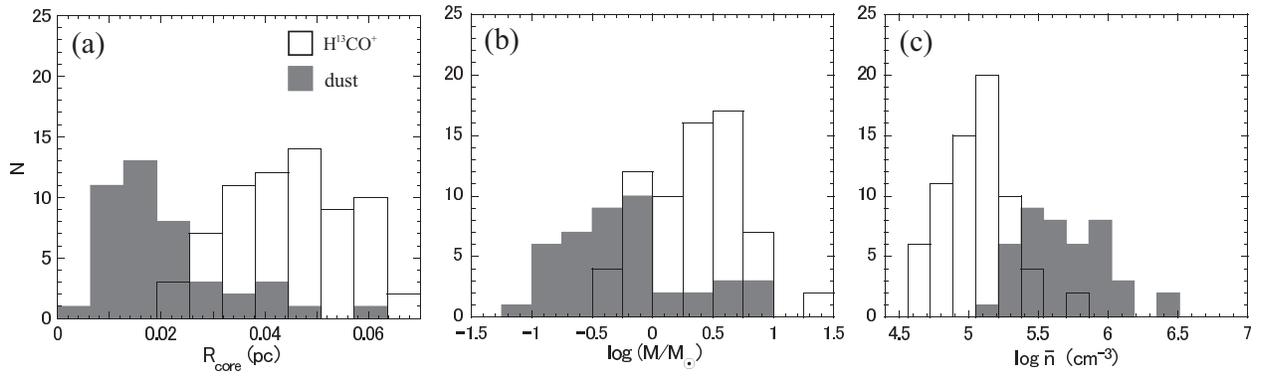}
\caption{Histograms of (a) the radius, (b) core mass, 
(c) mean density of  the H$^{13}$CO$^+$ cores 
({\it open histograms}) and  the 850 $\mu$m dust cores
({\it grey histograms}) in the $\rho$ Ophiuchi main cloud. 
}  
\label{fig:h13co+coreproperty}
\end{figure}

In Table \ref{tab:rhooph}, we present the physical properties of the 68
identified cores. 
In Table \ref{tab:h13co+core}, we summarize the minimum, maximum,
mean, and median values of each quantity.
The histograms of the radius ($R_{\rm core}$), LTE mass ($M_{\rm LTE}$),
and mean density ($\bar{n}$) of the H$^{13}$CO$^{+}$ cores 
are shown in Figure \ref{fig:h13co+coreproperty}. 
For comparison, the radius, mass, and mean density 
of the 850 $\mu$m cores listed in Table 7 of \citet{jorgensen08}
are indicated by the grey histograms in Figure \ref{fig:h13co+coreproperty}. 
We selected only the 850 $\mu$m cores located 
inside the same observed area as our H$^{13}$CO$^+$ map.
We also removed cores whose radii are smaller than the SCUBA beam size
of 14'' and corrected the radii of the other cores for the telescope beam size 
by using eq. [2] of \citet{ikeda07}. 
In this procedure, 43 dust cores were selected using 
the 2$\sigma$ noise level of 0.04 Jy beam$^{-1}$.
The minimum, maximum, mean, and median values of each quantity of 
the 850 $\mu$m cores are summarized in Table 
\ref{tab:scubacore}.
%fn
We note that as mentioned in \S \ref{sec:global}, in the 850 $\mu$m data 
the structures larger than 1.5' ($\approx 0.06$ pc) are removed during
the data reduction process and therefore 
the number of dust cores with radii larger than about 0.03 pc 
is likely to be underestimated.
%f

The radius of the H$^{13}$CO$^+$ cores ranges from 0.022 to 0.069pc and 
its distribution has a single peak at around the mean of 0.045pc 
(see Figure \ref{fig:h13co+coreproperty}a).
The maximum radius of the H$^{13}$CO$^+$ cores is only about three times 
as large as the minimum radius. In other words, most of the cores
have similar size.
On the other hand, the distribution of the LTE mass is somewhat 
broad compared with that of the core radius. 
The LTE mass of the H$^{13}$CO$^+$ cores ranges from 0.4 to 22 $M_\odot$
and its distribution has a single peak at around the mean of 
3.3$M_\odot$ (see Figure \ref{fig:h13co+coreproperty}b).  
The maximum mass is about 56 times as large as the minimum.
The mean density of the H$^{13}$CO$^+$ cores also shows 
a similar broad distribution, ranging 
from $3.7\times10^4$ to $5.8\times10^5$ cm$^{-3}$, 
with a single peak at around the mean density of $1.4\times 10^5$
cm$^{-3}$ (see Figure \ref{fig:h13co+coreproperty}c).

These distributions of the physical quantities of the 
H$^{13}$CO$^+$ cores are qualitatively similar to
those of the 850 $\mu$m cores, although the mean radius
and mass of the H$^{13}$CO$^+$ cores are about twice those 
of the 850 $\mu$m cores. 
On the other hand, the mean density of the H$^{13}$CO$^+$ cores is 
 smaller than that of the 850 $\mu$m cores by a factor of 5. 
This large difference in the mean density probably comes from 
the fact that the 850 $\mu$m emission tends to trace the higher
 density parts of the H$^{13}$CO $^+$ cores 
\citep[see][]{ikeda09}.
Another reason is due to the overlap effect. 
It is likely that dense cores often overlap one another along the line
of sight. Although such overlapped cores can be separated in some
degree for the H$^{13}$CO$^+$ cores because of their
velocity information, such separation is impossible
for the two-dimensional clumpfind method that was used 
for the 850 $\mu$m  map.
In fact, the distribution of the core radius for the 850 $\mu$m
cores has a tail toward the larger radius. 
A more prominent tail can be seen in the distribution of the 
core mass shown in Figure \ref{fig:h13co+coreproperty}b, 
where the mean core mass
is about 2.6 times as large as the median.
The tail in the core mass, however, can be caused by the 
temperature difference. For all the dust cores, 
the uniform temperature of $T=15$ K is assumed
in \citet{jorgensen08}. 
For several massive dust cores, however, the temperatures
may be underestimated because they are located 
in the Oph A region where the temperatures
are likely to be higher because of active star formation.

\begin{figure}
\plotone{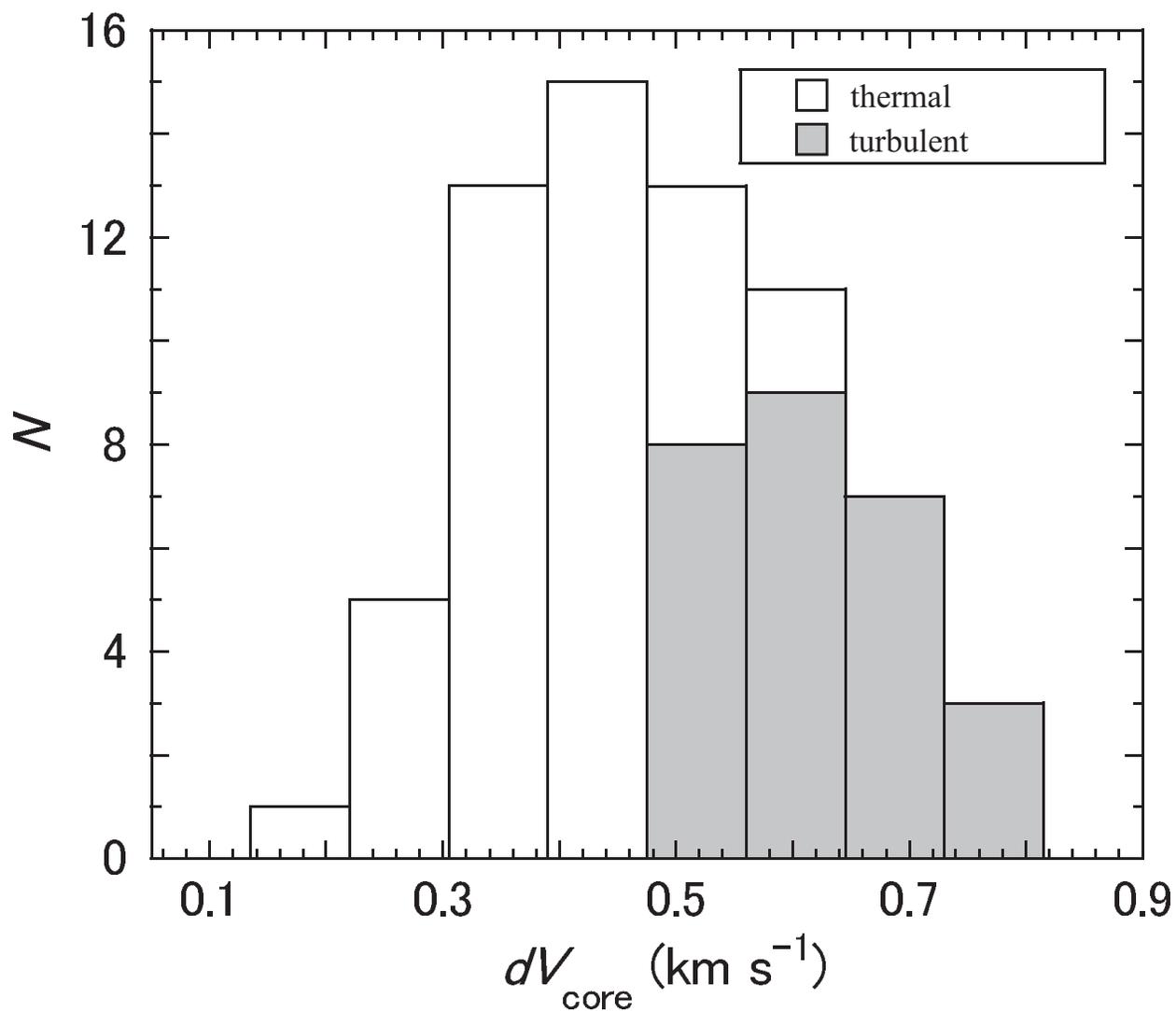}
\caption{Histogram of the FWHM line width of the H$^{13}$CO$^+$ cores 
in the $\rho$ Ophiuchi main cloud. The open and grey histogams indicate
 the thermal and turbulent cores, respectively.
Note that the gas temperature is assumed to be 12 K for all the cores
 except for those located in the Oph A region, for which 
$T=18$ K.
}
\label{fig:velocitywidth}
\end{figure}

The distribution of the FWHM line width of the H$^{13}$CO$^+$ cores
are shown in Figure \ref{fig:velocitywidth}. 
The line width ranges from 0.19 to 0.77 km s$^{-1}$ with 
a mean of 0.49 km s$^{-1}$.
Following \citet{myers91}, we classified our cores into two groups:
``thermal core'' and ``turbulent core'', 
on the basis of the critical line width, $dv_{\rm cr}$, expressed by 
\begin{equation}
dv_{\rm cr}=\left[8\ln2 \, k_B T 
\left({1\over \mu m_H}+{1\over m_{\rm obs}}\right)\right]^{1/2} \ . 
\label{eq:velocity width}
\end{equation}
The critical line width of H$^{13}$CO$^+$ is estimated 
to be 0.50 and 0.62 km s$^{-1}$ for $T=12$ K and 18 K, respectively.
Figure \ref{fig:velocitywidth} 
indicates that about 40 \% of the cores are classified as
the turbulent core.
However, almost all the turbulent cores have transonic turbulent motions,
and highly turbulent cores ($dv_{\rm core} \gtrsim 1$ km s$^{-1}$) 
found in massive star forming regions such as Orion A \citep{ikeda07} 
are not seen in the $\rho$ Ophiuchi main cloud.

\subsection{Comparison Between Cores with and without YSOs}
\label{subsub:yso}

Recent observations of the Ophiuchi molecular cloud complex with the
Spitzer Space Telescope \citep{padgett08} allow 
us to compare the physical properties  of H$^{13}$CO$^+$ cores 
associated with and without YSOs \citep{jorgensen08}. 
Previous studies have suggested that the physical properties of dense
cores associated with YSOs are different from 
those of cores without YSOs 
\citep[e.g.,][]{benson89,jijina99,walsh07}. 
Based on the spectral indices at the near- and mid-infrared wavelengths,
 \citet{jorgensen08} classified YSOs in L1688 into 
4 classes that represent the following evolutionary stages of YSOs: 
Class I, Flat Spectrum, Class II, and Class III.  
We use their YSO catalog to identify the H$^{13}$CO$^+$ cores with YSOs.
We note that in their classification, the Class 0 YSOs, the youngest
objects, are included in Class I. 
For example, the prototypical Class 0 object, VLA 1623, is
classified as Class I in their list.

\begin{figure}
\plotone{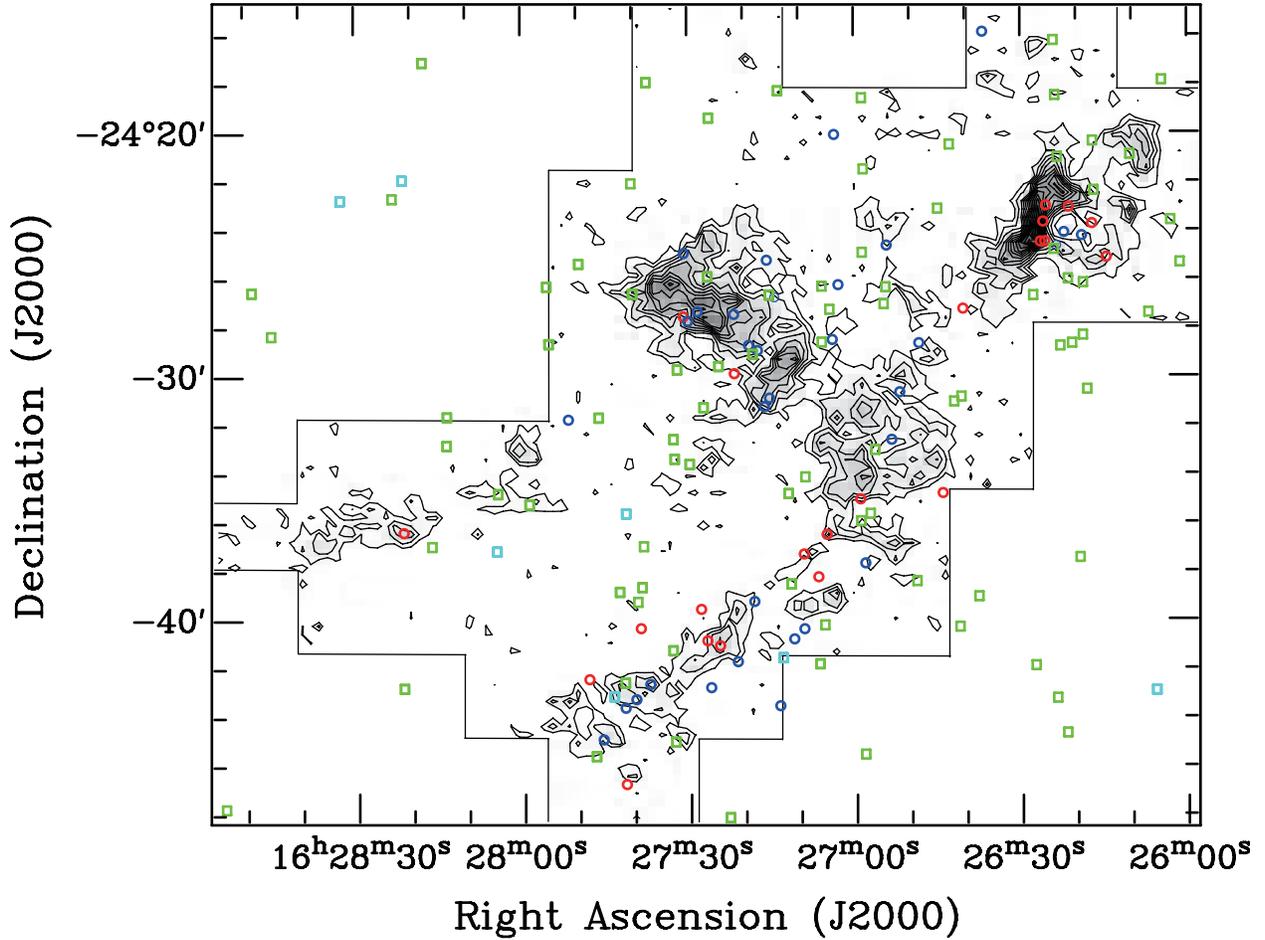}
\caption{Distribution of YSOs identified with the Spitzer Space
 telescople. Contours indicate the total integrated intensity of 
the H$^{13}$CO$^+$ ($J=1-0$) emission, as shown in 
Figure \ref{fig:global}a.
The red and blue circles are for Class I sources including 
Class 0 and Flat Spectrum sources, respectively. 
The green and cyan squares are
 for Class II and Class III  sources, respectively.
The cores associated with the circles are defined as protostellar cores
in this study.}
\label{fig:stardistribution}
\end{figure}

The spatial distribution of the YSOs is shown in Figure 
\ref{fig:stardistribution}, which 
indicates that the distribution of the Class I and Flat Spectrum 
sources follows that of the H$^{13}$CO$^+$ ($J=1-0$) emission. 
On the other hand, more evolved YSOs, the Class II and III sources, 
tend to be distributed in the entire observed area 
with no strong correlation with the distribution of 
the H$^{13}$CO$^+$ ($J=1-0$) emission.  
Therefore, we expect that the Class I and Flat Spectrum sources are 
probably embedded in the dense gas, and use 
the two classes of YSOs 
to divide the H$^{13}$CO cores into the following two groups: 
protostellar and starless cores.
If either a Class I or Flat Spectrum object is located within 
the extent of a H$^{13}$CO$^+$ core, 
we regard the core as a protostellar core.
The others are classified as starless cores, 
even if Class II or III sources are located within the cores.
In this classification, a Class I or Flat Spectrum source is sometimes 
located within the extents of two or more cores which 
partly overlap one another along the line of sight.
We classified all such cores as protostellar cores
because we have no information on the line of sight velocities
of the YSOs. Therefore, the number of protostellar cores
is likely to be overestimated.
According to the above procedure, we found 34 protostellar 
 and 34 starless cores.

\begin{figure}
\plotone{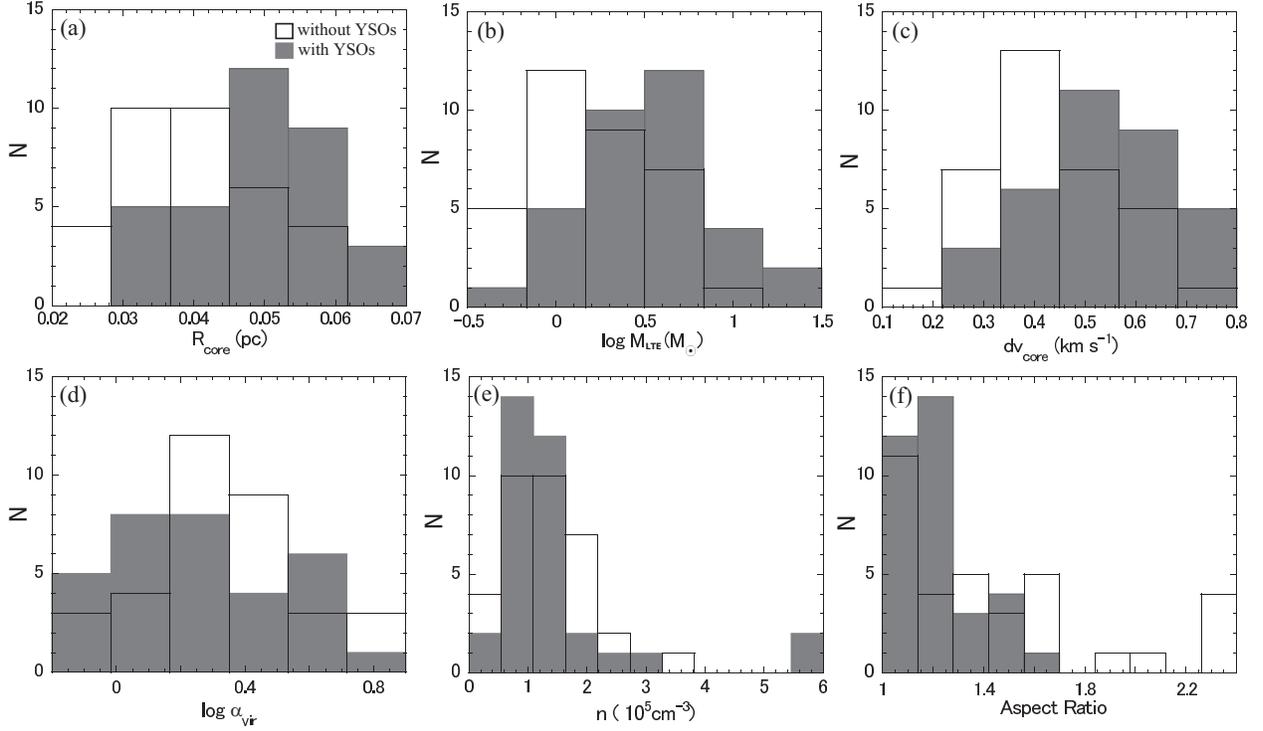}
\caption{Histograms of (a) the radius, 
(b) LTE mass, (c) FWHM line width, (d) virial ratio,
(e) mean density, and (f) aspect ratio of 
the H$^{13}$CO$^+$ cores.
For each panel, the open and grey histograms  are for
the cores without and with YSOs, respectively.
}
\label{fig:ysohistogram}
\end{figure}

Figure \ref{fig:ysohistogram} compares the physical properties of 
the starless and protostellar cores. 
The minimum, maximum, mean, and median values are summarized in
Table \ref{tab:yso}. The radii (Figures \ref{fig:ysohistogram}a) 
and masses (Figures \ref{fig:ysohistogram}b) of the protostellar cores
tend to be larger than those of the starless cores.
In particular, the mean LTE mass of the protostellar cores is 
about twice as large as that of the starless cores.
These results are consistent with those of the N$_2$H$^+$ cores in
NGC1333 \citep{walsh07} and the H$^{13}$CO$^+$ cores in Orion A
\citep{ikeda07}.

The line widths of the protostellar cores tend to be larger  
than those of the starless cores (Figures \ref{fig:ysohistogram}c).
A similar tendency is seen in the 
NH$_3$ cores observed by \citet{benson89} and 
\citet{jijina99},
who interpreted that stellar winds and
protostellar outflows create local turbulence.
If stellar winds and protostellar outflows enhance internal turbulent
motions for the protostellar cores, such cores must have had 
virial ratios smaller than the current values in the prestellar phase.
Although the distribution of the virial ratios for the protostellar cores
appears to be broader than that of the starless cores, 
there seems to be no significant difference 
in the virial ratio between the protostellar and starless cores
(Figures \ref{fig:ysohistogram}d):
for both the starless and protostellar cores, the virial ratio has 
a mean of about $2-3$, ranging from 0.4 to 8.
Here, the dimensionless parameter $a$ in 
eq.[\ref{eq:virial mass}] is adopted to be unity, 
the value of a uniform sphere, 
for both the starless and protostellar cores.
If for protostellar cores, the value of a centrally 
condensed density distribution, $a=5/3$, is adopted, 
the virial ratios of the protostellar cores tend to be somewhat 
smaller than those of the starless cores that 
are expected to have flattened density distributions.
This appears to contradict the idea that outflows enhance 
the virial ratios of the parent cores.
About 60 \% of the cores have virial ratios smaller than 2, for 
which the net internal energies are negative.
Therefore, we conclude that the majority of the cores are 
likely to be self-gravitating.
%For the  starless cores, 3 (9\%) and 16 (47\%) out of 34
%have virial ratios smaller than $\alpha_{\rm vir}=1$ and 2, respectively.
%For the protostellar cores, 7 (20\%) and 22 (65\%) out of 34
%have virial ratios smaller than $\alpha_{\rm vir}=1$ and 2,
%respectively.
%Therefore, we conclude that the majority of the cores (56\%) are 
%likely to be self-gravitating because the net internal energies
%are negative, i.e., $\alpha_{\rm vir}\le 2$. 

Figure \ref{fig:ysohistogram}e shows that the distributions 
of the mean densities of the protostellar and starless cores 
 are similar to each other, although some of the protostellar
cores have very large densities.
Figure \ref{fig:ysohistogram}f indicates
that the starless cores tend to have larger aspect ratios
than those of the protostellar cores. 
For example, for the starless cores, 5 (15\%) out of 34 have 
aspect ratios larger than 2, whereas for the protostellar cores, 
no cores have such large aspect ratios.
The starless cores with aspect ratios larger than 2 
have large virial ratios ($\alpha_{\rm vir} \gtrsim 3$)
and are likely to be gravitationally unbound.

\subsection{Correlation among Physical Properties of Dense Cores}
\label{subsec:correlation}

\subsubsection{Line Width-Radius and Mass-Radius Relations}

The line width measured from molecular line emission provides us with
information on kinetic properties such as thermal motion and turbulence.
Based on data in the literature, \citet{larson81} found that 
the line width of molecular clouds and cores is well correlated with 
the radius and mass, and the correlations are approximately of power-law form. 
He suggested that the power-law relations
 stem from processes of interstellar turbulence cascade. 
The strong correlations of the power-law relations also 
suggest that the molecular clouds and cores are nearly in virial equilibrium.
Since his pioneering work, many studies have been carried out 
to investigate the dependence of the line width on its radius
for various molecular clouds and cores using different molecular
 emission lines \citep[e.g.,][]{sanders85,dame86,heyer04}.
Observations based  mainly on low-density tracers 
such as $^{12}$CO ($J=1-0$) and $^{13}$CO ($J=1-0$) 
 have generally supported Larson's idea
\citep[e.g.,][]{myers83,fuller92}. 
In contrast, observations based on higher-density tracers such as
C$^{18}$O, CS, and H$^{13}$CO$^+$ have often shown 
that the dependence of the line width on its radius
tends to be very weak 
\citep[e.g.,][]{tachihara02,ikeda07,ikeda09,saito08}.
For massive star forming regions, the shallower power-law relations
have also been found \citep{caselli95, plume97}.

\begin{figure}
\epsscale{1.0}
\plotone{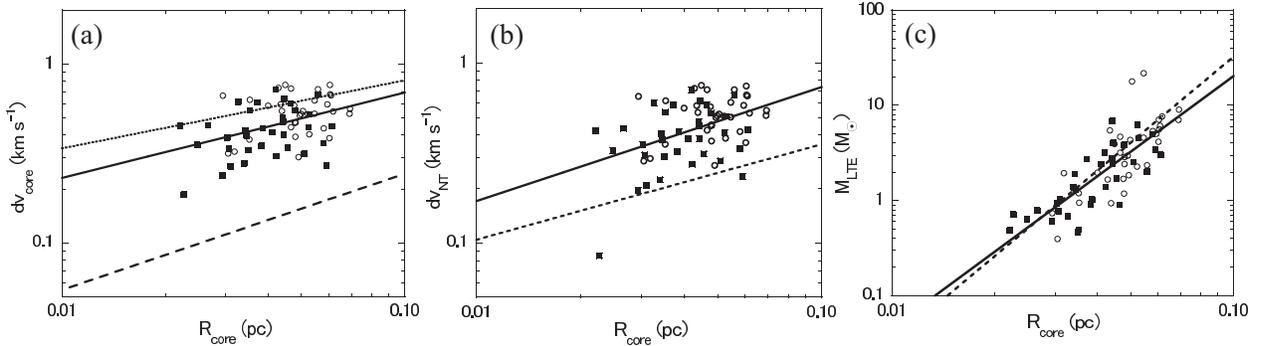}
\caption{
(a) Line width-radius relation, (b) Nonthermal line width-radius
 relation, and (c) Mass-radius relation of the H$^{13}$CO$^+$ cores.
The filled squares and open circles are for the cores without and with
 YSOs, respectively.
In each panel, the best-fit power-law is indicated by the 
solid line. In panel (a) the Larson and 
Heyer \& Brunt  relations are plotted by 
the dotted and dashed lines, respectively.
Note that we extrapolated these relations down to 0.01 pc, for comparison.
In panels (b) and (c), the dashed lines indicate the Caselli \& Myers relation 
and the relation of $M_{\rm LTE}=4\pi \mu m_H \bar{n}R_{\rm core}^3/3 \ 
(\propto R_{\rm core}^3)$, respectively.
}
\label{fig:velocity width}
\end{figure}

The line width-radius relation for the
identified cores is presented in Figure \ref{fig:velocity width}a.
The filled squares and open circles indicate the starless 
and protostellar cores, respectively.
As mentioned in \S \ref{subsub:yso}, the protostellar cores tend to 
have larger line widths and larger radii
than those of the starless cores. 
Such a tendency can be recognized in Figure \ref{fig:velocity width}a.
The best-fit power-law for all the cores is given by 
\begin{equation}
\log ({dv_{\rm core}/{\rm km \ s^{-1}}}) =
(0.319\pm 0.179)+ (0.478\pm 0.131) 
\log ({R_{\rm core}/{\rm pc}})  \ ,
\end{equation}
with ${\cal R}=0.410$. 
As representative examples, the line width-radius relations derived by
\citet{larson81} and \citet{heyer04} are also plotted 
in Figure \ref{fig:velocity width}a with
the dotted and dashed lines, respectively.
\citet{heyer04} applied the Principal Component Analysis to 
derive the line width-radius relation inside molecular clouds
from $^{12}$CO observations.
The line widths of the H$^{13}$CO$^+$ cores tend to be
significantly larger than those of the Heyer \& Brunt relation
and somewhat smaller than those of the Larson relation.
We note that the line widths of the Larson relation 
tend to be larger over the scale 0.01 $-$ 0.1 pc compared to 
the line width-radius relations derived by other authors
\citep[e.g.,][]{solomon87,myers83}
and thus may correspond to the upper bound of the relations.
Although the power index of our cores is not far from those of 
\citet{larson81} and \citet{heyer04}, the correlation is very weak.
Such a weak correlation is also seen for the H$^{13}$CO$^+$ cores 
in Orion A \citep{ikeda07}.
In \S \ref{sec:discussion} we will discuss the line width-radius 
relation of the H$^{13}$CO$^+$ cores in more detail.

To separate the turbulent motion from the thermal motion
within each core, we plot in Figure \ref{fig:velocity width}b 
the nonthermal line width $dv_{\rm NT} [\equiv (dv_{\rm tot}^2-8\ln 2 \,
k_B T/\mu m_H)^{1/2}]$ against core radius.
The best-fit power-law for all the cores,
shown by the solid line in Figure \ref{fig:velocity width}b, is 
given by
\begin{equation}
\log (dv_{\rm NT}/{\rm km \ s}^{-1}) = (0.501\pm 0.218)+ (0.634\pm 0.159)
\log (R/{\rm pc})
\end{equation}
with ${\cal R}=0.440$.
For comparison,  the nonthermal line width-radius relation 
for low-mass cores compiled by \citet{caselli95}, 
%%%$(dv_{\rm NT}/{\rm km \ s}^{-1}) = 1.51(R/{\rm pc})^{0.53}$, 
$(dv_{\rm NT}/{\rm km \ s}^{-1}) = 1.2(R/{\rm pc})^{0.53}$, 
is indicated by the dashed line, 
%fn
where we replaced the radius of \citet{caselli95} ($R_{CM}$) by 
the radius of our definition. See the Appendix in detail. 
%f
Although our relation has a power-law index similar to
that of \citet{caselli95}, the coefficient is somewhat larger.
This suggests that turbulence within our cores is not 
as much dissipated as the Caselli \& Myers relation predicts.
This tendency is different from that of 
the H$^{13}$CO$^+$ cores in Orion A 
 where the $dv_{\rm NT}$-$R_{\rm core}$ relation
agrees well with the Caselli \& Myers relation 
\citep{ikeda07}.
In \S \ref{sec:discussion} we will compare our cores with 
the Orion A cores and discuss this point in more detail.

%fn
Recently, \citet{volgenau06} attempted to measure the relationship
between the line width and size inside three protostellar cores
having multiple protostars in the Perseus molecular cloud,
using the Regions-Of-Contrast (ROC) analysis proposed by 
\citet{ostriker01}. 
Our line width-radius relation is essentially different from
that derived from the ROC analysis, and therefore, it is 
difficult to directly compare the two relations.
In the ROC analysis, the size is just a region size that is not related 
to a dense core.
When the ROC analyis is applied to substructures inside a core, 
the line width-size relation tends to have a large scatter, 
especially when the objects have complicated structures 
like those observed by \citet{volgenau06}, in which 
the disk rotation,  disk instability, and 
stellar feedback such as stellar winds and outflows 
influence the local structures significantly.
In contrast, in our analysis, the size (or radius) 
corresponds to that of an identified core.
%(Note that the distribution of the H$^{13}$CO$^+$ emission is localized
%in the data space, unlike $^{12}$CO and $^{13}$CO emision, and therefore
%the identified structures are expected to reasonably trace the real 
%structures in molecular clouds.)  
Therefore, the relatively large scatter in our line width-radius relation is 
likely to reflect the difference in the dynamical states of the cores
(see also \S \ref{subsec:virial analysis}).
In fact, recent numerical simulations of turbulent molecular clouds
demonstrated that the cores identified directly from the
3D density distribution are in various dynamical states and 
show similar large scatter in the line 
width-radius relation \citep{nakamura08,offner08}.
%f

In Figure \ref{fig:velocity width}c, we plot the LTE mass 
against core radius.
The best-fit power-law is given by 
\begin{equation}
\log (M_{\rm LTE}/M_\odot)=(3.95\pm 0.34) +(2.64\pm 0.25) \log
(R_{\rm core}/{\rm pc})
\end{equation}
where ${\cal R}=0.790$. For comparison, the relation
of $M_{\rm LTE}=4\pi \mu m_H \bar{n} R_{\rm core}^3 /3$ is indicated by 
the dashed line in Figure \ref{fig:velocity width}c, where
$\bar{n}=1.35\times 10^5$ cm$^{-3}$.
The power-law index of $\sim 3$ indicates that 
the mean density of each core is comparable to the
critical density of H$^{13}$CO$^+$ ($J=1-0$), 
$n_{\rm cr}\simeq 8\times 10^4$ cm$^{-3}$.
The coefficient of our $M_{\rm LTE}$-$R_{\rm core}$ relation
is an order of magnitude larger than that of \citet{larson81}, which is
mainly based on $^{13}$CO ($J=1-0$) 
whose critical density is about $10^3$ cm$^{-3}$.
This tendency is in good agreement with that of the Orion A cores
\citep{ikeda07,ikeda09}.

\subsubsection{Virial Ratio}

\begin{figure}
\plotone{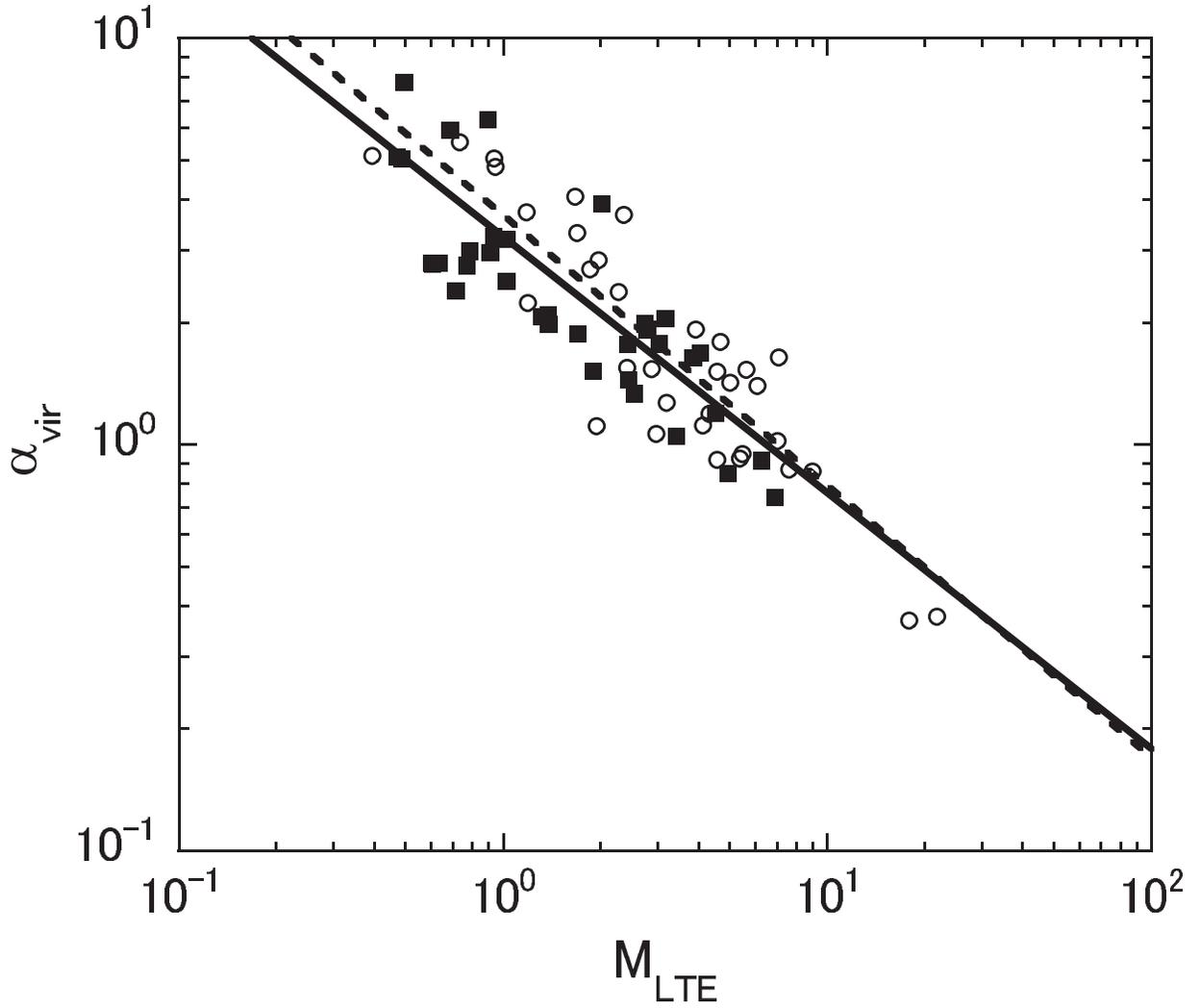}
\caption{
Virial ratio-mass relation
 of the H$^{13}$CO$^+$ cores.
The filled squares and open circles are the same as those in Figure
 \ref{fig:velocity width}.
The virial ratio of a self-gravitating core confined by ambient 
pressure is indicated by the dashed line \citep{bertoldi92}.
}
\label{fig:virialratio}
\end{figure}

The boundedness of a core is often estimated using the virial ratio.
In Figure \ref{fig:virialratio}, 
we plot against LTE mass the virial ratio calculated using
eq.[\ref{eq:virial mass}], where for both the protostellar and 
starless cores 
a dimensionless parameter $a$ is set to unity, the value of 
a uniform sphere.
The best-fit power-law of our cores is given by
\begin{equation}
%\log \alpha_{vir} =  (0.114\pm 0.045) + (-0.488\pm 0.086)
%\log (M_{\rm LTE}/M_\odot)
\log \alpha_{\rm vir} =  (0.528\pm 0.018) + (-0.600\pm 0.057)
\log (M_{\rm LTE}/M_\odot)
%\log \alpha_{\rm vir} =  (0.375\pm 0.024) + (-0.558\pm 0.075)
%\log (M_{\rm LTE}/M_\odot)
\end{equation}
with ${\cal R}=0.817$.
The virial ratio tends to increase with decreasing the LTE mass,
and has a mean of 2.4, ranging from 0.4 to 8. 
More than a half the cores have virial ratios
smaller than 2 and thus the majority of the cores are 
more or less self-gravitating.
According to \citet{bertoldi92}, a self-gravitating core confined by 
ambient pressure has a virial ratio of 
$\alpha_{\rm vir} = 2.06(M_{\rm LTE}/M_J)^{-2/3}$,
where $M_J$ is the Jeans mass defined by eq. [2.13] of
\citet{bertoldi92}.
For comparison, the virial ratio of a self-gravitating core
confined by ambient pressure is plotted by the dashed line in Figure
\ref{fig:virialratio}. 
Here, using the velocity dispersion of $0.35$ km s$^{-1}$ and the
ambient gas density of $10^5$ cm$^{-3}$ (see the derivation of these values 
in \S \ref{sec:discussion}), 
the Jeans mass is estimated to be about $2.4$ M$_\odot$,
almost equal to the median mass of the identified cores.
The best-fit power-law is very close to 
the virial ratio of a self-gravitating core confined by ambient
pressure.
The virial ratios of the identified cores
tend to be, however, smaller than that of a non self-gravitating, 
pressure-confined core 
with $\alpha_{\rm vir} = 2.9(M_{\rm LTE}/M_J)^{-2/3}$,
for a given core mass \citep{bertoldi92}. 
Therefore, for our cores, both the self-gravity 
and ambient pressure play an important role 
in dynamics of the cores.
In \S \ref{sec:discussion} we will discuss the dynamical states 
of our cores in more detail.

\section{Discussion}
\label{sec:discussion}

\subsection{Core Mass Spectrum in the $\rho$ Ophiuchi Main Cloud}

\begin{figure}
\plotone{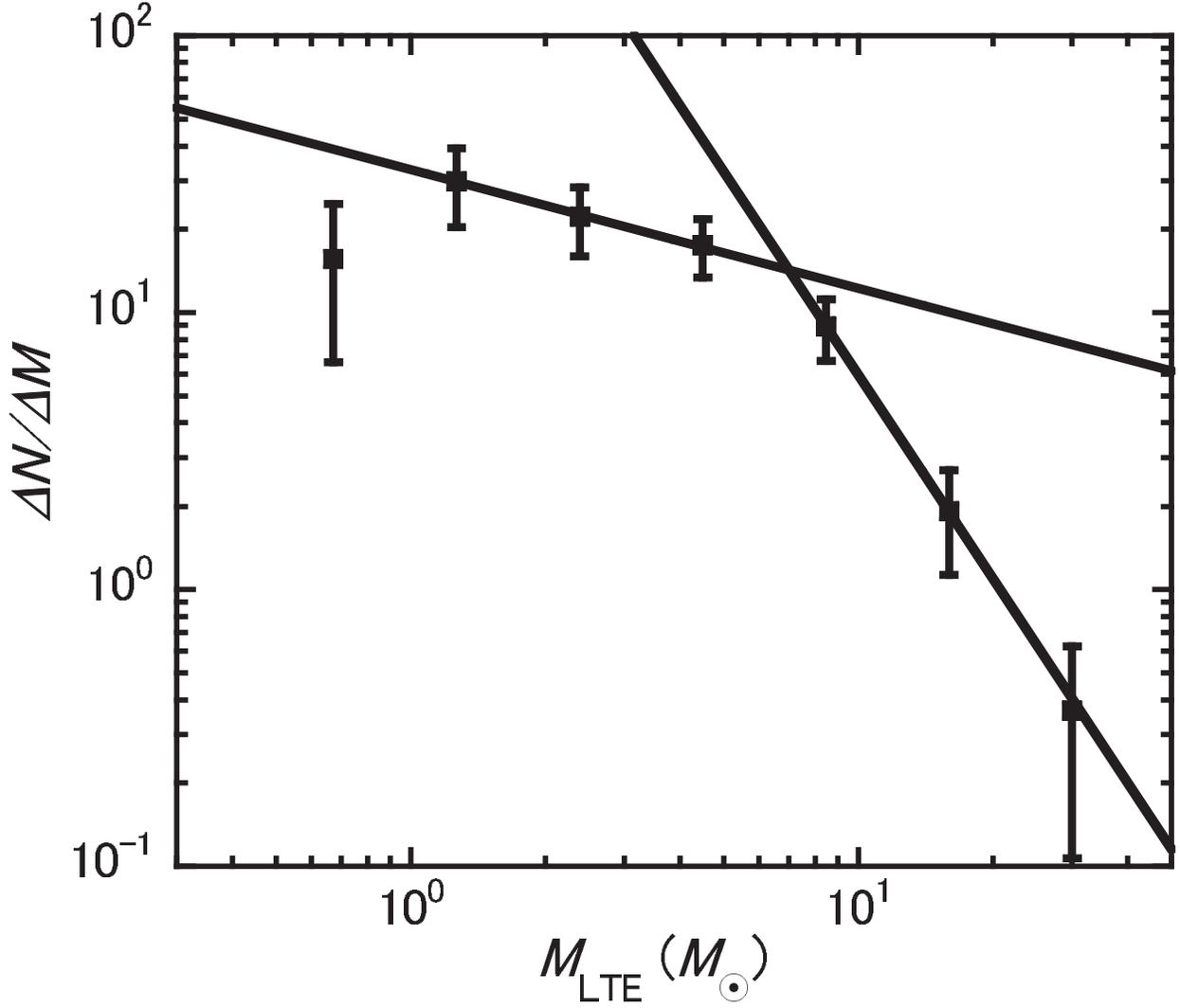}
\caption{Mass spectrum of  the H$^{13}$CO$^+$ cores 
in the $\rho$ Ophiuchi main cloud.
The error bars correspond to $\sqrt{N}$ counting statistics,
where $N$ is the number of cores in each mass bin.
The solid lines are the least square fits 
in the interval of $1M_\odot \lesssim M_{\rm LTE} \lesssim 7M_\odot$
($\Delta N/\Delta M \propto M_{\rm LTE}^{-0.43}$) 
and $M_{\rm LTE}\gtrsim 7 M_\odot$ 
($\Delta N/\Delta M \propto M_{\rm LTE}^{-2.4}$).
}
\label{fig:cmf}
\end{figure}

The mass spectrum of our cores is plotted in Figure 
\ref{fig:cmf}.
The core mass spectrum appears to be fitted by a two-component power-law.
There seems to be a break at around $7M_\odot$, 
about twice the mean LTE mass of 3.4$M_\odot$.
Above the break, the mass spectrum is steeper,
while below the break, it is flattened.  
The mass spectrum can be fitted by 
the following two-component power-law:
\begin{eqnarray}
dN/dM &\propto& M^{-0.43 \pm 0.03} \ {\rm for} \ 
M_{\rm LTE} \lesssim 7M_\odot \\
 &\propto& M^{-2.4 \pm 0.3} \ {\rm for} \ M_{\rm LTE} \gtrsim 7M_\odot  \ .
\label{eq:cmf}
\end{eqnarray}
This mass spectrum is broadly consistent with
those of the dust cores in $\rho$ Oph, 
although for the dust cores the low-mass parts are 
somewhat steeper and the break masses are smaller
\citep{motte98,johnstone00,stanke06,stamatellos07,enoch08,simpson08}.
For example, \citet{stamatellos07} revised the mass spectrum of
the dust cores identified by \citet{motte98} by reestimating 
the dust temperatures. The revised mass spectrum is fitted by 
a two-component power-law with a low-mass index of $-1.5$, 
a high-mass index of $-2$, and a break at around 1$M_\odot$.
\citet{simpson08} reanalyzed the SCUBA 850 $\mu$m archive data 
and obtained the core mass spectrum that can be fitted by a three-component
power-law with a low-mass index of $-0.7$, intermediate-mass index
of $-1.3$, high-mass index of $-2.35$, and two breaks at around
0.7$M_\odot$ and 2$M_\odot$.
We note that the slopes and the break masses of the core mass spectra
are affected by the number of bins, if it is not appropriately set
%\citep{ballesteros02,schneider04,rosolowsky05,smith08}.
\citep{rosolowsky05}.
Thus, the cumulative form of the core mass spectra
is sometimes used when the number of cores is small,
although it requires a more complicated uncertainty analysis
\citep[see e.g.,][]{reid06}.
As for the parameters of the clumpfind, the slopes 
are almost independent of the threshold and stepsize of the clumpfind 
as long as they are appropriately set \citep{ikeda09b}.
For the core mass spectra mentioned above, the difference in the break
mass is likely to come from the fact that 
the H$^{13}$CO$^+$ emission traces more extended, less dense structures 
than the dust continuum emissions. 
The break mass may also be affected by
the ``confusion'' which means the situation that 
multiple cores with similar $v_{\rm LSR}$ are overlapped along the same 
line of sight and therefore cannot be separated into individual cores
using the clumpfind \citep[see][]{ikeda07,ikeda09}.

Based on the ISOCAM observations, \citet{bontemps01} identified over 200
 YSOs in $\rho$ Oph and derived 
the mass function of 123 Class II YSOs that 
is well fitted by a two-component power-law 
with a low-mass index of $-1.15$, a high-mass index of $-2.7$, and a
 break at around 0.55$M_\odot$.
The mass spectrum of the H$^{13}$CO$^+$ cores is 
roughly similar in shape to the stellar IMF in $\rho$ Oph,
although the slopes in the low mass and high mass parts 
of the H$^{13}$CO$^+$ core mass spectrum are somewhat shallower 
and the break mass is one order of magnitude larger.

\subsection{Dynamical State of H$^{13}$CO$^+$ Cores in the $\rho$ Ophiuchi Main
  Cloud}
\label{subsec:virial analysis}

Virial theorem is useful for analyzing the dynamical states of dense
cores.
The virial equation for a uniform spherical core is given by
\begin{equation}
{1 \over 2} \frac{\partial ^2 I}{\partial t^2} = U + W + S
\end{equation}
where the terms, $I$, $U$, $W$, and $S$, denote
the moment of inertia, internal kinetic energy, 
gravitational energy including the effect of 
magnetic field, and surface pressure, respectively, and
are given as follows \citep{nakano98}:
\begin{eqnarray}
%U&=&  {3k_B T M \over \mu m_H}  \\ 
%K&=&{3M dv_{\rm NT}^2\over 8\ln 2} \\ 
U&=&{3M dv_{\rm tot}^2\over 8\ln 2} \\ 
W&=&-{3\over 5}{GM^2\over R_{\rm core}} 
\left[1-\left({\Phi \over \Phi_{\rm cr}}\right)^2 \right] \\
S&=& -4 \pi R^3 P_{\rm ex} \ .
\end{eqnarray}
The values $\Phi$ and $\Phi_{\rm cr}$ are, respectively, the magnetic flux 
penetrating the core and the critical magnetic flux above which 
the magnetic field can support the core against the self-gravity. 
Here, the core is assumed to have condensed from much lower 
density medium and the radius of the magnetic flux tube
penetrating the core is much smaller than that before contraction
[see \S 2 of \citet{nakano98}].
$P_{\rm ex}$ is the surface pressure including both thermal 
and turbulent components.
$M$ is the core mass and is chosen to be the LTE mass of the core.

All the above terms in the virial equation except 
$P_{\rm ex}$ and $\Phi$ can be estimated from the
physical quantities listed in Table \ref{tab:rhooph}.
It is difficult to estimate the surface pressures exerted on
individual cores directly from our data.
Instead of deriving the surface pressures of individual cores,
we adopt a representative surface pressure 
from the average densities and velocity dispersions 
that were measured in subclupms Oph A, B, C, E, and F:
 $\left<P_{\rm ex}\right> \approx \left<\rho\right>  \left<\sigma\right>^2$.
Here, $\left<\rho\right>$ is the average density 
and $\left<\sigma\right>$ is 
the average velocity dispersion including both the thermal and turbulent 
components as 
$\left<\sigma\right>^2=\left<\sigma_{\rm NT}\right>^2+c_s^2 $,
where $\left<\sigma_{\rm NT}\right>$ and $c_s$ are the 
velocity dispersion of the nonthermal component 
and the isothermal sound speed, respectively. 
The average density and velocity dispersion were evaluated
as $\left<\rho\right>/(2.33m_H)\approx (0.5-1) \times 10^5 $ cm$^{-3}$ and 
$\left<\sigma\right> \approx 0.3-0.4$ km s$^{-1}$, respectively.
The average surface pressure is then given by 
$\left<P_{\rm ex}\right>/k_B \approx
3 \times 10^6$ K cm$^{-3}$. This value is about a few times as large as 
the thermal pressure [$\approx 10^5$ cm$^{-3} \times (12-18)$ K 
$\approx (1-2) \times 10^6$ K cm$^{-3}$].
In the following, we adopt the above $\left<P_{\rm ex}\right>$
 as a representative value for our cores.
This surface pressure is about a half the average internal
pressure inside the cores ($\approx 6\times 10^6$ K cm$^{-3}$) and 
 close to the lower limit of the critical ambient pressure 
for the dust cores obtained by \citet{johnstone00}.

On the other hand, there is only one reliable measurement of the magnetic 
field strength toward the $\rho$ Ophiuchi main cloud.
\citet{crutcher93} performed OH Zeeman effect measurements toward 
two positions in $\rho$ Oph with a beam size 18'.
For one of the two positions that well covers the entire 
observed area of our
data, they derived the line-of-sight magnetic field strength 
of about 10 $\mu$G at the low density cloud envelope of 
$N_{\rm H_2} \approx 5\times 10^{21}$ cm$^{-3}$ \citep[see also][]{troland96}.
Based on this measurement, \citet{crutcher99} estimated the magnetic 
flux normalized to the critical value to be 
$\Phi/\Phi_{\rm cr}\simeq 0.4$.
If we adopt this value as a representative value of $\Phi/\Phi_{\rm cr}$
for our cores, the magnetic effect is likely to be minor: it reduces 
the gravitational energy term only by 16\%.
However, recent turbulent simulations have demonstrated that cores 
formed out of turbulent clouds can have much larger values of 
$\Phi/\Phi_{\rm cr}$ than the cloud initial values \citep[e.g.,][]{dib07} 
and thus it is very difficult to assess
the values of $\Phi/\Phi_{\rm cr}$ for the individual cores
without direct measurements of the magnetic fields associated with the cores.
In the following, we simply assume $\Phi/\Phi_{\rm cr}=0$.

\begin{figure}
\plotone{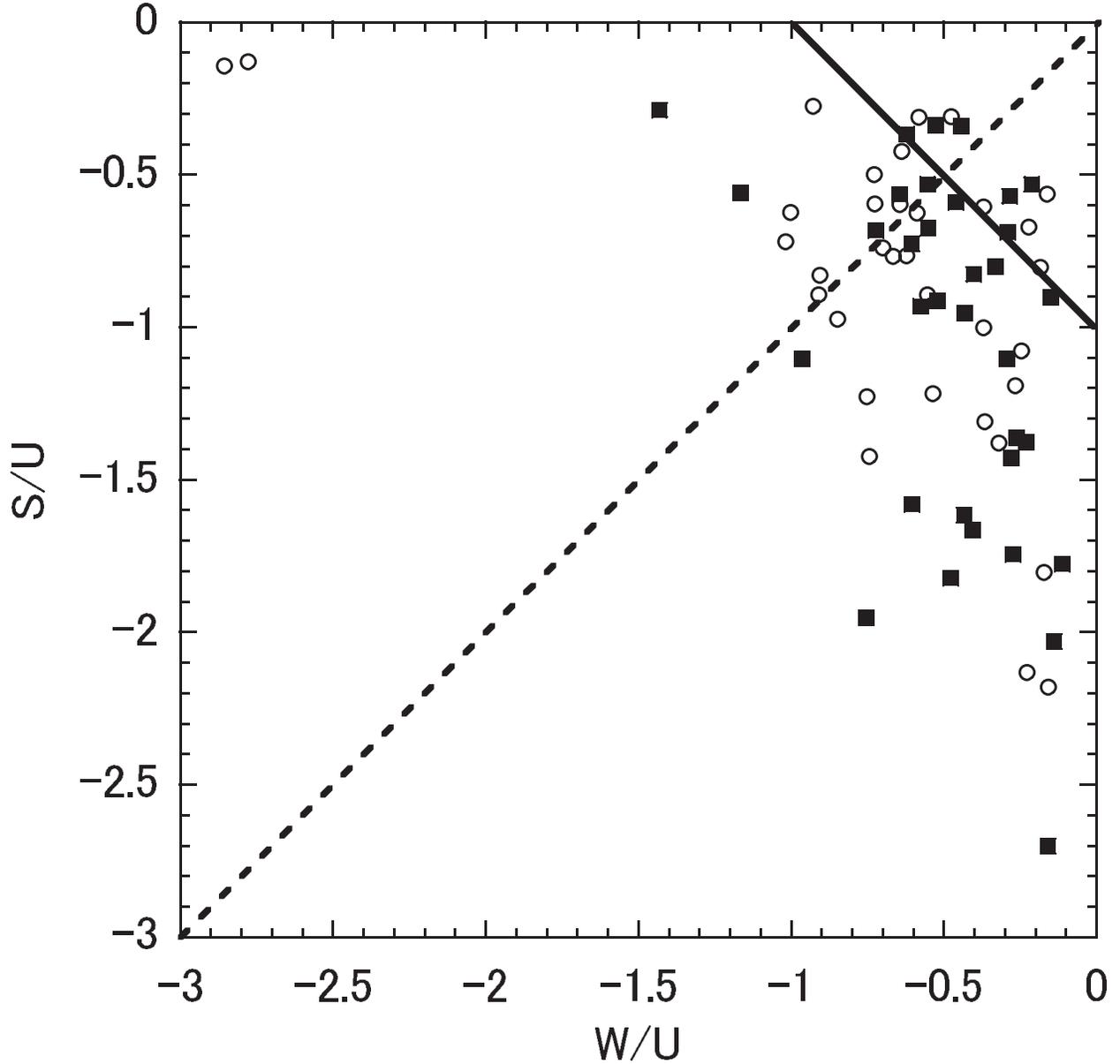}
\caption{
Relationship between the surface term, $S$, and the gravitational term,
 $W$, in the virial equation. They are normalized to the internal
 kinetic energy term, $U$ [Note that $W/U = -(\alpha_{\rm vir})^{-1}$].
The solid line indicates the virial equilibrium,
 $U+W+S=0$. The dashed line indicates the line at which $W=S$.
For the cores that lie below the solid line, the value of $U+W+S$ is
 negative and thus expected to be bound.  All  others are 
unbound and expected to disperse away, if they do not gain
 more mass through accretion and/or merging with other cores, or reduce
 internal support through turbulence dissipation.
The filled squares and open circles are the same as those in Figure
 \ref{fig:velocity width}.
}
\label{fig:virial}
\end{figure}

The equilibrium line, $U+W+S=0$, 
is shown by the solid line in Fig.~\ref{fig:virial}, where 
the surface term ($S$) is plotted against 
the gravitational energy term ($W$), 
both normalized to the internal energy term ($U$).
For the cores that lie below the solid line, the value of $U+W+S$ is
 negative and thus expected to be bound.  
All the others are unbound and expected to disperse away, if they do not gain
 more mass through accretion and/or merging with other cores, or reduce
 internal support through turbulence dissipation.
More than a half the cores lie below the equilibrium line, and are thus bound: 
they are expected to collapse and form stars.
This is also true even in the presence of magnetic field as long as 
the magnetic flux does not exceed about a half the critical value.
Furthermore, the majority of the cores lie 
below the line of $S=W$ (dashed line), 
indicating that 
the surface term is more important than the
gravitational energy term.
Even for the cores lying above the line $S=W$, the surface pressure 
appears to be dynamically important for almost all such cores
because the deviation from the line is small.
We note that this plot should be valid statistically, but
for any individual points the true surface pressures could be higher or
lower than the representative surface pressure.

Our virial analysis indicates that the formation and evolution of dense cores 
in the cluster environment are likely to be strongly influenced by the
external compression due to local turbulent motions 
(see also Dobashi et al. 2001 for importance of external pressures
 at cloud scales).
Such local compression may be responsible for the formation
of binary and multiple stars or formation of substellar objects 
(brown dwarfs or planetary mass objects) in each core 
under the cluster environment \citep{hennebelle03,gomez07,whitworth07}.

\subsection{Comparison with Orion A}

The Orion A molecular cloud is the nearest giant molecular cloud
located at a distance of 480 pc \citep{genzel81}, about 4 times as distant as 
 the $\rho$ Ophiuchi main cloud. 
Recently, \citet{ikeda07} carried out the H$^{13}$CO$^+$ ($J=1-0$) core
survey in the whole region of Orion A using  
the Nobeyama 45 m telescope and identified 236  cores 
by the same method as adopted in the present paper.
Their core sample is ideal for comparison with our cores
because the same molecular emission line, 
the same telescope, and the same core identification procedure
are used.
In this subsection, we compare the physical properties of our 
 cores in $\rho$ Oph with those in Orion A.

Figures \ref{fig:orion1} and \ref{fig:orion3} 
compare the physical properties of the $\rho$ Oph cores 
with those of the Orion A cores.
The open squares and crosses indicate the $\rho$ Oph cores 
 and the Orion A cores, respectively.
The mean radius of the $\rho$ Oph cores, 0.045 pc, 
is about three times smaller than that of the Orion A cores, 0.14 pc.
This probably reflects the different spatial
resolutions (or different distances) for both the observations. 
In fact, the area where the $\rho$ Oph cores are distributed
is well separated from that of the Orion A cores on the plots
of the line width-radius and mass-radius relations
(see Figure \ref{fig:orion1}).
On the other hand, the mean mass of the $\rho$ Oph cores,
3.35 $M_\odot$, is only about 3.6 times smaller than that 
of the Orion A cores, 12 $M_\odot$. 
This is because the mean density of the Orion A
cores, $1.6\times 10^4$ cm$^{-3}$, is much smaller than that of 
the $\rho$ Oph cores, $1.4\times 10^5$ cm$^{-3}$.
In particular, the mean density of the Orion A cores is smaller than 
the critical density of the H$^{13}$CO$^+$ ($J=1-0$),
$8\times 10^4$ cm$^{-3}$, and therefore 
the Orion A cores are likely to contain substructures that could not be
 spatially resolved.

The mean line width of the $\rho$ Oph cores, 0.488 km s$^{-1}$, 
is almost the same as that of the Orion A cores, 0.52 km s$^{-1}$,
in spite of the different core radii.
The ranges of the line widths are also similar to each other
(see Figure \ref{fig:orion1}). On the other hand, 
the virial ratio of the Orion A cores seems three times larger than 
that of the $\rho$ Oph cores for a given LTE mass, as shown in 
Figure \ref{fig:orion3}.  
This difference may be explained also by the different spatial
resolutions 
as follows. The virial ratio is proportional to both the radius 
and the square of the line width for a given LTE mass. 
Since the line widths of both the cores are comparable to each other 
and the radii of the Orion A cores are roughly three times larger than 
those of the $\rho$ Oph cores, the virial ratio of Orion A becomes three
times 
larger than that of $\rho$ Oph for a given LTE mass.

To elucidate whether or not the different spatial resolutions 
dominate the different core properties in the two clouds, 
we try to eliminate the effect
of the different spatial resolutions as follows.
We smoothed the original $\rho$ Oph data on a coarser grid 
with a grid spacing of 80" by convolving with a 2D Gaussian kernel 
with a FWHM of 80" in the $\alpha$-$\delta$ space. 
The smoothed data correspond to the data that would be observed 
by the Nobeyama 45 m telescope 
if the $\rho$ Ophiuchi main cloud were 
located at the same distance as the Orion A molecular cloud.
Then, applying the clumpfind to the smoothed data, we reidentified 
16 dense cores, which are plotted by the filled circles in 
Figures \ref{fig:orion1} and \ref{fig:orion3}. 
In Figures \ref{fig:orion1}a and \ref{fig:orion1}b,
the $\rho$ Oph cores identified from the smoothed data are located just in
the areas where the Orion A cores are distributed. 
In contrast, the virial ratios of the reidentified $\rho$ Oph cores 
 are somewhat
smaller than those of  Orion A for a given LTE mass
(see Figure \ref{fig:orion3}a).
The larger virial ratios for the Orion A cores may reflect 
the  larger $X_{\rm H^{13}CO^+}$ adopted by \citet{ikeda07}, 
$4.8\times 10^{-11}$, which is derived using the $^{13}$CO abundance 
measured in the cloud envelope of Taurus, which has much lower 
column densities than Orion A.
If we use the $^{13}$CO abundance obtained by \citet{frerking82}
for $\rho$ Oph, which has column densities comparable to 
the region observed by \citet{ikeda07}, 
the H$^{13}$CO$^+$ abundance is reduced by a factor of 2, close 
to our value.
Note that our value is in agreement with that of \citet{ikeda07} within
uncertainty mentioned by them. 
Using our value of $X_{\rm H^{13}CO^+}$, the distribution of the 
Orion A cores in the virial 
ratio-LTE mass plot becomes consistent with that of the $\rho$ 
Oph cores identified from the smoothed data, as presented
in Figure \ref{fig:orion3}b.
Although in that case, the LTE masses of the Orion A cores shift upward
in Figure \ref{fig:orion1}b, the distributions of 
the $\rho$ Oph and Orion A cores in the LTE mass-radius relation 
plot are still well overlapped with each other. 
Therefore, we conclude that there are no clear differences in the core
properties between the $\rho$ Oph and Orion A clouds.
The apparent differences in the core properties 
between the two  clouds
are likely to be caused mainly by the different spatial resolutions.

\subsection{Line Widths of the H$^{13}$CO$^+$ cores and 
Implication for Turbulent Generation}
\label{subsec:small scale turbulence}

Figure \ref{fig:orion1} shows that 
 there are no highly turbulent cores having 
$dv_{\rm core} \gtrsim 1$ km s$^{-1}$ in $\rho$ Oph.
According to \citet{ikeda07}, all the highly turbulent cores 
in Orion A are located within 1 pc from the M42 HII region 
and are probably influenced greatly by the nearby OB stars.
The turbulent cores are expected to be responsible for massive star 
formation \citep[see the discussion of \S 4.3 of][]{ikeda07}.
The absence of such highly turbulent cores in 
$\rho$ Oph may imply that 
massive O stars will not form in  $\rho$ Oph 
under the current environment.

It is also interesting that  the range of the line widths of the $\rho$
Oph cores identified from the smoothed data almost coincide 
with that of the original $\rho$ Oph cores. 
If we fit a single power-law to both the $\rho$ Oph cores
identified from the original and smoothed data, the best-fit result
becomes  $dv_{\rm core} = (0.599\pm 0.104)R_{\rm core}^{0.073\pm 0.060}$
with a very small correlation coefficient of 0.137, which is
shown by the solid line in Figure  \ref{fig:orion1}.
In other words, the line width appears to be almost 
independent of the core radius.
Since the cores identified from the smoothed data contain 
adjacent several cores identified from the original data, 
the nearly independent relationship between the line width and core
radius, or the nearly flat  line width-radius relation, 
suggests that the inter-core motions among the neighboring cores are 
almost comparable to the internal motions 
in the individual cores.
Such a feature is pointed out by \citet{andre07} who measured 
the velocity difference among the neighboring cores 
from the centroid velocities of the cores observed by N$_2$H$^+$ 
($J=1-0$).
If the line width-radius relation is flat 
($dv_{\rm core} \approx$ const.) and the core mass 
is proportional to $R_{\rm core}^3$, then the 
virial ratio is scaled as 
$\alpha_{\rm vir}\propto R_{\rm core}dv_{\rm core}^2/(GM_{\rm LTE})
\propto M_{\rm LTE}^{-2/3}$, a similar power-law to
that derived by \citet{bertoldi92}.
This again suggests the importance of ambient turbulent 
pressure in dynamics of the cores 
as discussed in \S \ref{subsec:virial analysis}.

The line width-radius relation of the $\rho$ Oph cores
 appears to be inconsistent with
one of the most reliable measurements of the line width-size relation, 
recently obtained by \citet{heyer04},
who found a strong correlation between the line width and size 
based on the $^{12}$CO observations toward 27 nearby giant molecular
clouds, despite the large differences in cloud environments
and local star formation activity. 
Their relation, plotted by the dashed line in 
Figure \ref{fig:orion1} 
[$(dv/{\rm km s^{-1}})= 1.1 (R/{\rm pc})^{0.65}$], 
leads to the interpretation that most of the turbulent energy 
comes from the largest scale that is comparable to the cloud size
(see the Appendix).
\citet{heyer07} claimed that the 
driving mechanisms of turbulence at small scales 
such as protostellar outflows and stellar winds may
not play an important role in dynamics of molecular clouds.
The FWHM line width ($\Delta V \sim 1.5 $ km s$^{-1}$)
and radius ($R\sim 1.5$ pc) measured 
from the $^{13}$CO ($J=1-0$) line toward the whole $\rho$ Ophiuchi 
main cloud are in good agreement with the Heyer \& Brunt relation.
In contrast, the $\rho$ Oph cores identified from the original data
tend to deviate upwards from the Heyer \& Brunt relation.
The nearly flat line width-radius relation of the $\rho$ Oph cores 
may also suggest that turbulent energy has been injected at the scales
smaller than $R\sim 0.35$ pc, at which our best-fit power-law
intersects with the Heyer \& Brunt relation.
This scale is in reasonable agreement with the characteristic length scale 
 of the outflow-driven turbulence estimated from the
 theoretical consideration \citep{matzner07,nakamura07}.
Based on $^{13}$CO and C$^{18}$O observations, \citet{swift08} 
derived a similar characteristic scale
of the outflow-driven turbulence of $\approx 0.05$ pc toward L1551.
However, the universality of the Heyer \& Brunt relation 
implies that the outflow-driven turbulence may be limited 
to the localized dense regions where active star formation occurs.
A caveat is that it is unclear whether the Heyer \& Brunt relation,
based on the $^{12}$CO observations, is still 
applicable to dense gas where stars are forming.
Therefore, the deviation of the line widths for $\rho$ Oph
from the Heyer-Brunt relation may be apparent.
However, it is worth noting that the nearly flat line width-radius 
relation is consistent with the virial ratio-mass relation of 
the pressure-confined, self-gravitating cores 
($\alpha_{\rm vir} \propto M^{-2/3}$).
This may suggest that the characteristic length scale of 0.35 pc
is related to the formation of dense self-gravitating structures.
%Even in such a  case, the additional injection of  turbulent
%motions may be needed to reproduce the nearly flat 
%line width-radius relation.
In any case, if such a small scale driving of turbulence is common in the
cluster environment, star formation in clusters
is likely to continue over several dynamical times.

On the other hand, the line widths of our cores deviate downward from the
Larson relation which may correspond to the upper bound 
of the observed line width-radius relations.
Until now, many authors have derived the line width-radius relations
for various molecular clouds and cores.
Those studies show a scatter in the index and  coefficient
of the power-law line width-radius relations.
The line widths of the Larson relation,
$(\Delta v/{\rm km s^{-1}}) = 1.9 (R/{\rm pc})^{0.38}$, tend to be somewhat 
larger over the range of  0.01 $-$ 0.1 pc compared to the other 
relations and thus may correspond to the upper bound of the  relations.
For example, on the basis of the Massachusetts-Stony Brook Galactic Plane
Survey data, \citet{solomon87} derived the relation of 
$(\Delta v/{\rm km s^{-1}}) = 2.1 (R/{\rm pc})^{0.5}$, not far from the 
Heyer \& Brunt relation.
In nearby low-mass star
forming regions \citet{fuller92} compiled the data of dense cores 
and derived the power-law relation of 
$(\Delta v/{\rm km s^{-1}}) = 1.0 (R/{\rm pc})^{0.4}$
for starless cores.
If the line width-radius relation of the low-density gas in $\rho$ Oph
follows the Larson relation,
instead of the Heyer \& Brunt relation, the smaller line widths 
of the H$^{13}$CO$^+$ cores may suggest that the
 cores have formed preferentially in regions where supersonic
turbulence  dissipated and thus the  cores
so formed have smaller line widths 
compared to the Larson relation.
If this is the case, the nearly independent relationship 
between the line width and core radius implies 
that the turbulent field at the small scales  has not been relaxed yet,
and therefore the timescale of core formation may be of 
the order of a dynamical time.
In that case, it is unclear 
how protostellar outflows contribute to maintain 
 supersonic turbulence in the cluster environment.

\section{Summary}
\label{sec:summary}

Using the archive data of the H$^{13}$CO$^+$ ($J=1-0$) molecular line
emission taken with the Nobeyama 45 m radio telescope, 
we analyzed the molecular gas distribution 
in the central dense region of the $\rho$ Ophiuchi main cloud.
We summarize the primary results of the paper as follows:

1. We compared the global distribution of the H$^{13}$CO$^+$ emission
with that of the 850 $\mu$m dust continuum emission and found 
that the overall spatial distributions are similar to each other, 
while the H$^{13}$CO$^+$ emission appears to cover a larger area than the 
dust continuum emission.
By comparing  between the H$^{13}$CO$^+$  and 850 $\mu$m
 maps, we revealed that the fractional abundance 
of H$^{13}$CO$^+$ relative to H$_2$ decreases with 
increasing the H$_2$ column density as 
$X_{\rm H^{13}CO^+}\propto N_{\rm H_2}^{-1/2}$. 
This tendency is consistent with the theoretical 
prediction \citep[e.g.,][]{nakano02}. 
The mean fractional abundance in the region is estimated 
to be $1.72 \times 10^{-11}$.

2. From the 3D data cube of the H$^{13}$CO$^+$ emission 
we identified 68 dense cores using the clumpfind method.
From comparison with the positions of YSOs recently 
identified by the Spitzer space telescope, 
the  cores are classified into the following two groups: 
34 protostellar and 34 starless cores.
The radii, masses, and line widths of the protostellar cores 
tend to be larger than those of the starless cores.
The virial ratio tends to increase with decreasing the LTE mass,
although we found no significant difference in the virial ratio 
between the protostellar and starless cores.
Furthermore, the virial ratios of the cores 
can be well described by the model of a self-gravitating
 core confined by ambient pressure derived by \citet{bertoldi92},
suggesting that both self-gravity and ambient pressure play 
an important role in dynamics of the cores.

3. The mass spectrum of the H$^{13}$CO$^+$ cores can be fitted 
by a two-component power-law that resembles the stellar IMF
in $\rho$ Oph. 
The core mass spectrum has a break at around $7
M_\odot$, close to the mean LTE mass of the cores.
It implies that the stellar IMF may be at least partly determined 
by the core mass distribution.

4. Applying the virial analysis, we conclude that most of the 
cores are bound and are expected to collapse.
Furthermore, for the majority of the cores, the surface pressure term
is more important than the gravitational energy term.
This result suggests that 
 in the cluster environment the formation and evolution of the 
 cores may be regulated largely by the surface pressures.

5. We compared the physical properties of the H$^{13}$CO$^+$ cores
in $\rho$ Oph with those in Orion A.
To eliminate the effect of the different distances, we smoothed the
original $\rho$ Oph data into a coarser grid
so that the spatial resolutions coincide with each other.
We identified 16 cores from the smoothed data using the clumpfind.
These cores would be observed 
by the Nobeyama 45 m telescope 
if the $\rho$ Ophiuchi main cloud were located at the same 
distance as the Orion A molecular cloud.
The physical properties of the $\rho$ Oph cores identified 
from the smoothed data appear to resemble those of Orion A.
The fact that the mean density of the Orion A cores 
($1.6\times 10^4$ cm$^{-3}$) is somewhat smaller than the critical 
density of H$^{13}$CO$^+$ ($J=1-0$) transition of  $8\times 10^4$ cm$^{-3}$
suggests that the Orion A cores contain  clumpy substructures that
could not be spatially resolved.

6. The range of the line widths of the $\rho$ Oph cores 
identified from the smoothed data almost coincides with that 
of the original $\rho$ Oph cores. 
This suggests that the line width-radius relation
is more or less flat over the range of 0.01 $-$ 0.1 pc
for the dense gas detected by H$^{13}$CO$^+$.
Such a flat relation is inconsistent with 
that derived from the $^{12}$CO observations by \citet{heyer04}
who found that the line width correlates strongly with
the size for the low-density molecular gas.
At the scales below 0.3 pc, the line widths of 
the $\rho$ Oph cores deviate upwards from the Heyer \& Brunt relation.
This may be due to turbulence driven by protostellar outflows.
However, if the actual line widths of the low-density gas in
$\rho$ Oph agree with the values of the Larson relation,  
then the nearly independent relationship between the line width and core
radius may be interpreted as evidence that the cores have 
formed preferentially in regions where supersonic turbulence dissipated.

\appendix
\section{Comparison with Other Line Width-Radius Relations}

In the present paper, the FWHM line width and core radius
defined by our clumpfind method are used for showing the line
width-radius relation of the identified cores.
Since the different authors use the different definitions of the line
width and core radius
\citep{heyer04,caselli95,larson81}, we here convert 
the other line width-radius relations using our definitions.

\citet{heyer04} applied the PCA analysis to derive 
the relationship between the line wdith and the size inside the 
molecular clouds. Their line width $\Delta v_{\rm HB}$ 
and size $L_{\rm HB}$ correspond to the autocorrelation widths on both 
the spatial and the velocity axes.  
In the following, we transfer them into our FWHM line width and radius
to do a direct comparison with our result.

If both a line and intensity profiles of a core 
can be approximated by Gaussian profiles, then the line width 
and size defined by \citet{heyer04} are given by 
$\Delta v_{\rm HB} = 2\sigma_v$ and 
$L_{\rm HB} = 2\sigma_L$, where
$\sigma_v$ and $\sigma_L$ are the dispersions of 
the line and intensity profiles, respectively
\citep[see][]{heyer04}.
On the other hand, the FWHM line width is defined
as $\Delta v_{\rm FWMH}=\sqrt{8\ln 2} \sigma_v$.
Thus, the FWMH line width can be rewritten as
$\Delta v_{\rm FWHM}=1.18\Delta v_{\rm HB}$.

In the present paper, the  core radius, $R_{\rm core}$, 
is determined by taking the projected area enclosed 
by the 2 $\sigma$ level contour and by computing the radius 
of the circle required to reproduce the area.
For a core with a Gaussian profile, our core radius
is given by 
$R_{\rm core} = L_{\rm HB} \sqrt{\ln (T_{\rm peak}/ T_{\rm th})}$,
where $T_{\rm peak}$ and $T_{\rm th}$ are the peak antenna temperature
and the threshold temperature of the 2 $\sigma$ level, respectively.
From Table \ref{tab:rhooph}, the mean peak antenna temperature of 
the 68 identified cores is estimated to be $\left<T_{\rm peak}\right>=1.14$K.
If we adopt $\left<T_{\rm peak}\right>$ as a representative value,
$R_{\rm core} = 0.91L_{\rm HB}$.

\begin{figure}
\epsscale{0.7}
\plotone{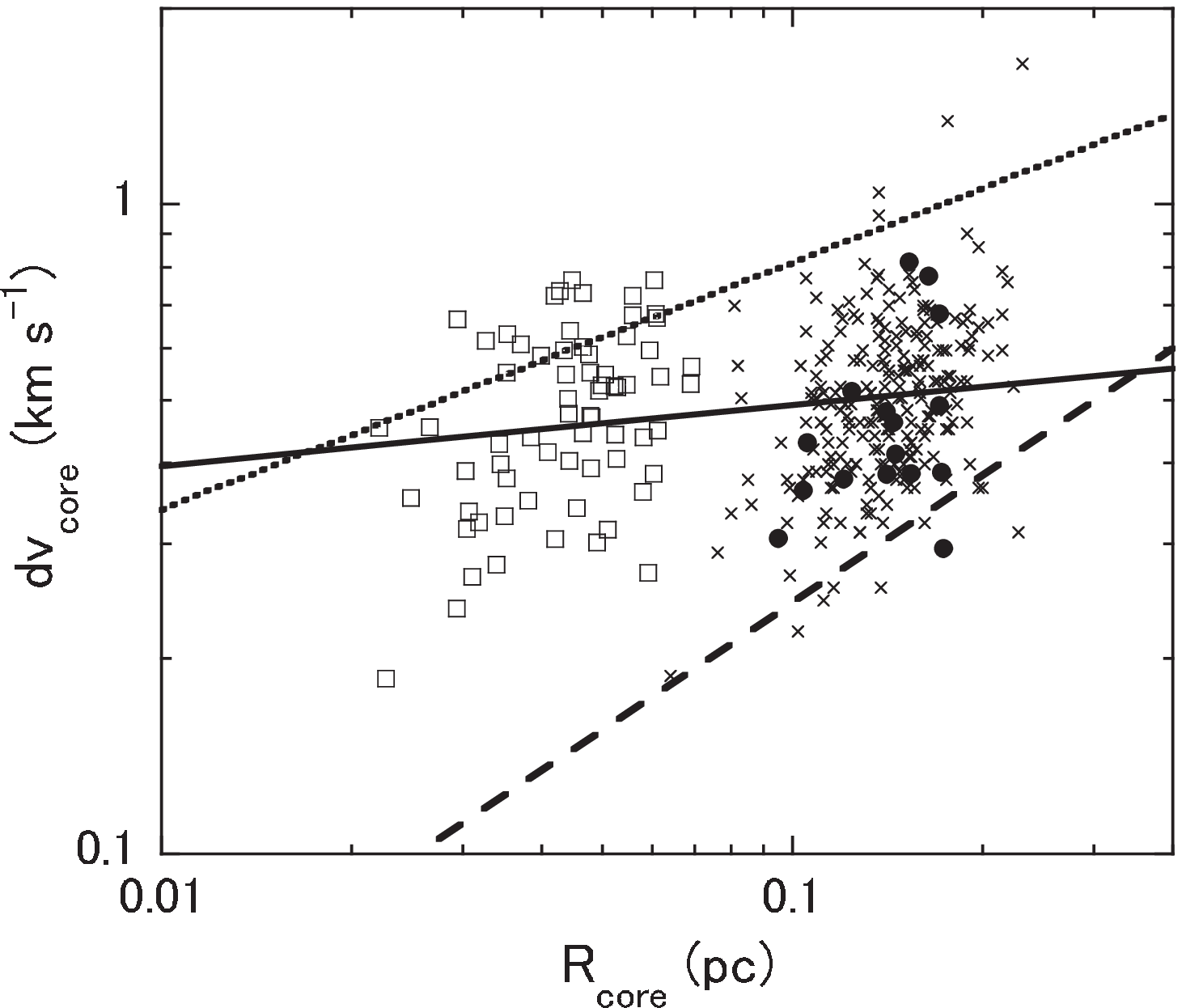}
%\plotone{figure/orion_dv-r.eps}
%\plotone{figure/orion_dv-m.eps}
\plotone{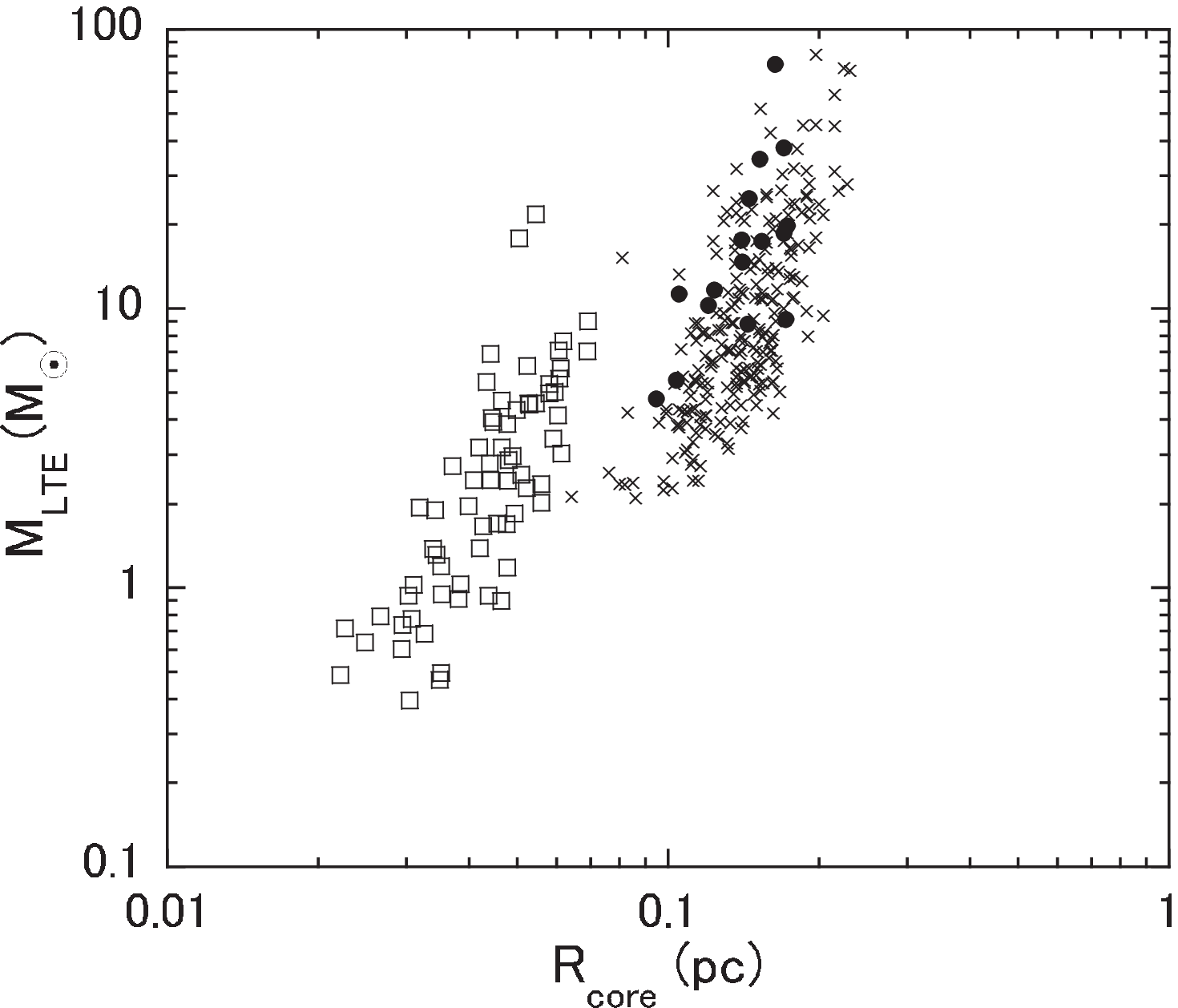}
\epsscale{1.0}
\caption{
Line width-radius relation ({\it upper panel}) 
and mass-radius relation ({\it lower panel})
of the H$^{13}$CO$^+$ cores
toward the $\rho$ Ophiuchi main cloud 
({\it open squares}) and toward the Orion A molecular cloud 
 ({\it crosses}).
The $\rho$ Oph cores identified from the smoothed
 data are indicated by filled circles, for comparison.
The solid line in the upper panel indicates the best-fit power-law
for the $\rho$ Oph cores identified  from both the original
and smoothed data. The dotted and dashed lines indicate the 
Larson and the Heyer \& Brunt relations, respectively. 
}
\label{fig:orion1}
\end{figure}

\begin{figure}
%\plotone{figure/new/orion_virial-mnew.eps}
\plottwo{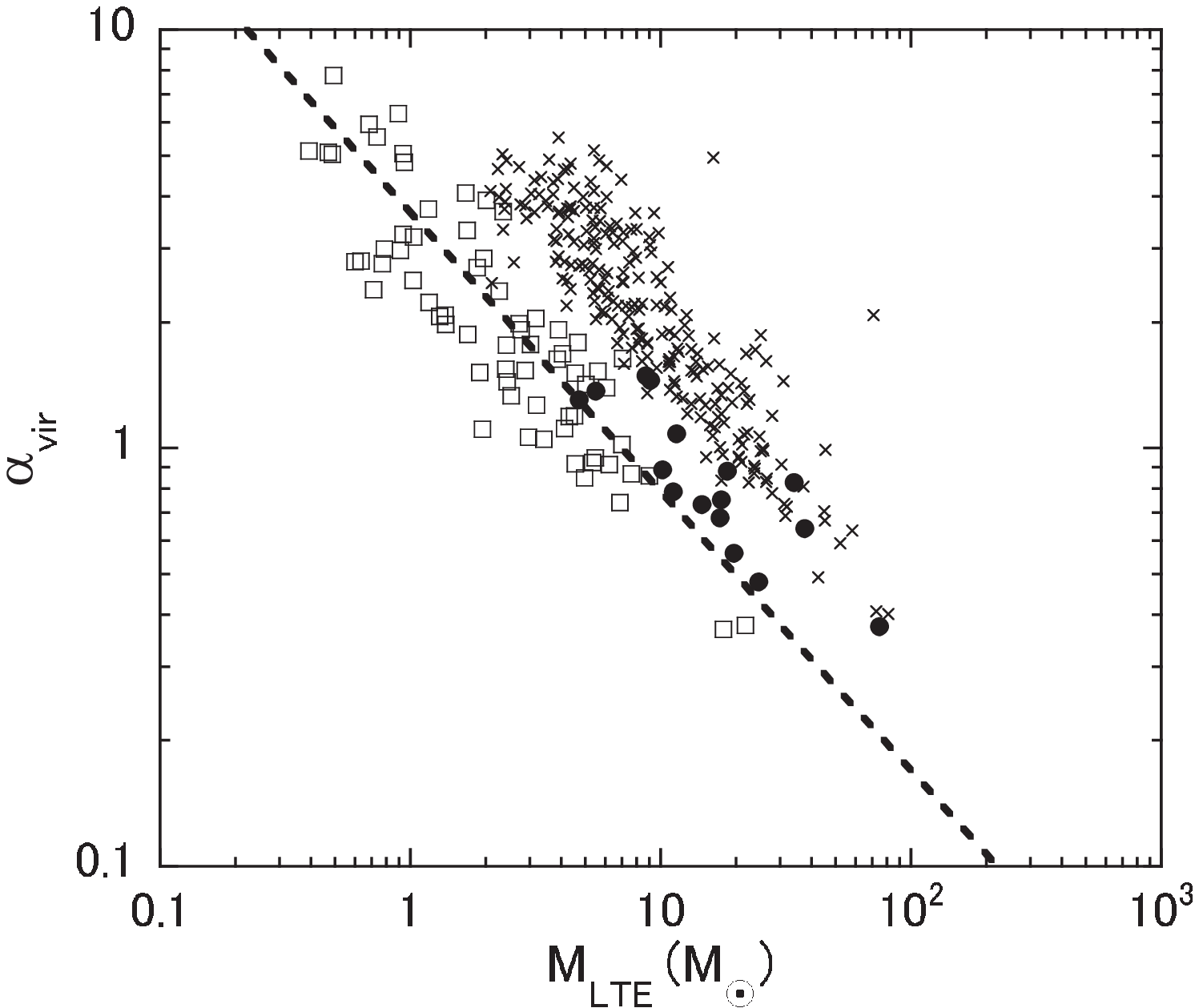}{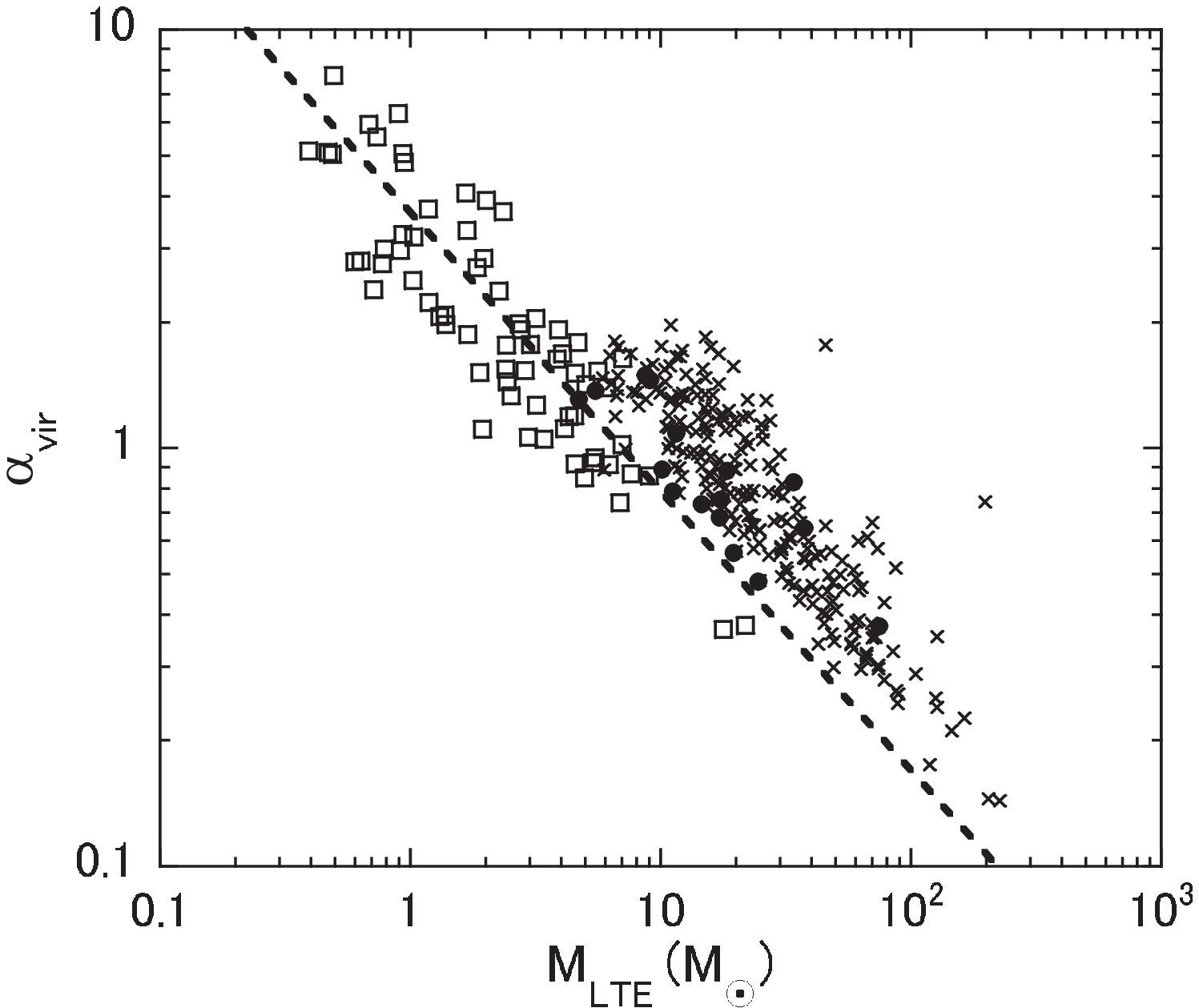}
\caption{
Virial ratio-mass relation
 of the $\rho$ Oph cores and the Orion A cores.
For Orion A, the fractional abundances of 
H$^{13}$CO$^+$ are set to 
$4.8\times 10^{-11}$ (adopted by Ikeda et al. 2007)
and $1.72\times 10^{-11}$
(adopted for $\rho$ Oph in the present paper) 
in panels (a) and (b), respectively.
The open squares, crosses, and filled circles are the same as those of
 Figure \ref{fig:orion1}.
The dashed lines are the same as that of Fig. \ref{fig:virialratio}.
Note that the Jeans mass given by eq. [2.13] of \citet{bertoldi92}
is almost the same for both the original 
and smoothed $\rho$ Oph cores.
}
\label{fig:orion3}
\end{figure}

According to \citet{heyer04}, the linewidth-size relation for GMCs
can be fitted by the power-law of 
$(\Delta v_{\rm HB}/{\rm km \ s}^{-1})=0.87 (L_{\rm HB}/{\rm pc})^{0.65}$.
Using $\Delta v_{\rm FWHM}=1.18\Delta v_{\rm HB}$ and 
$R_{\rm core} = 0.91L_{\rm HB}$, the Heyer \& Brunt relation 
can be rewritten as 
$(\Delta v_{\rm FWHM}/{\rm km \ s}^{-1}) = 1.09 (R_{\rm core}/{\rm pc})^{0.65}$,
which is shown by the dashed line in Figure \ref{fig:orion1}a.

\citet{caselli95} also used the different definition for the core
radius. They used the half-maximum contour to define the core size.
Assuming that a core has a Gaussian shape, the half-maximum radius
is equal to 
$R_{\rm CM}=\sqrt{\ln 2/\ln (T_{\rm peak}/T_{\rm th})} R_{\rm core}$,
and $R_{\rm CM}=0.65 R_{\rm core}$ for the mean peak antenna temperature
of $\left<T_{\rm peak}\right>=1.14$K.
We thus replace the original Caselli \& Myers relation, 
$(dv_{\rm NT}/{\rm km \ s}^{-1}) = 1.51(R_{\rm CM}/{\rm pc})^{0.53}$, 
with 
$(dv_{\rm NT}/{\rm km \ s}^{-1}) = 1.2(R_{\rm core}/{\rm pc})^{0.53}$.

\citet{larson81} used the 3D velocity dispersion ($\sigma_{\rm 3D}$) 
and the maximum core size ($L_{\rm max}$). 
Assuming that $L_{\rm max}=2R_{\rm core}$, his original relation,
$(\sigma_{\rm 3D}/{\rm km \ s}^{-1})=1.1(L_{\rm max}/{\rm pc})^{0.38})$,
is rewritten as 
$(\Delta v_{\rm FWHM}/{\rm km \ s}^{-1}) = 1.9 (R_{\rm core}/{\rm pc})^{0.38}$.

\acknowledgments 
This work is supported in part by a Grant-in-Aid for Scientific Research
of Japan (19204020, 20540228) and a Grant for Promotion of 
Niigata University Research Projects. 
We thank Jes J\o rgensen for kindly giving us the data of YSOs in L1688.
We also thank Zhi-Yun Li, Doug Johnstone, and Philippe Andr\'e for valuable comments,
and Chris Brunt and Mark Heyer for helpful comments 
that improved the presentation of \S 
\ref{subsec:small scale turbulence} and the Appendix.

\clearpage

%%%%%%%%%%%

%%%%%%%%%%%%%%%%%%%%%%%%%%%%%%%%%%%%%%%%%%%%%%%%%%%%%%
%%%%%%%%%%%%%%%%%%%%%%%%%%%%%%%%%%%%%%%%%%%%%%%%%%%%%%

\clearpage

\begin{deluxetable}{lllccccccccccc}
\rotate
\tablecolumns{14}
\tabletypesize{\footnotesize}
\tablewidth{0pt}
\tablecaption{Properties of the H$^{13}$CO$^+$ cores in the $\rho$ Ophiuchi
 main cloud
\label{tab:rhooph}}
\tablehead{
\colhead{ID} & 
\colhead{R.A.} & 
\colhead{Decl.} & 
\colhead{$v_{\rm LSR}$} & 
\colhead{$T^*_{A, \rm peak}$} & 
\colhead{$R_{\rm core}$} & 
%%%\colhead{$R_{\rm core}$\tablenotemark{d}} & 
\colhead{$R_{\rm core}$} & 
\colhead{Aspect} &
\colhead{$dv_{\rm core}$} & 
\colhead{$M_{\rm LTE}$} & 
\colhead{$M_{\rm vir}$} & 
\colhead{$\alpha_{\rm vir}$} &
\colhead{$\bar{n}$} & 
\colhead{YSOs\tablenotemark{a}}  \\
\colhead{} & 
\colhead{(J2000.0)} & 
\colhead{(J2000.0)} & 
\colhead{(km s$^{-1}$)} & 
\colhead{(K)} & 
\colhead{(arcsec)} & 
\colhead{(pc)} &
\colhead{ Ratio}  & 
\colhead{(km s$^{-1}$)} & 
\colhead{($M_\odot$)} & 
\colhead{($M_\odot$)} & 
\colhead{} &
\colhead{($10^5$ cm$^{-3}$)} & 
\colhead{} 
}
\startdata
1 & 16 26 7.3  & -24 20 27.6  & 3.15  & 1.50  & 72.9  & 0.0442  & 
 2.33 & 0.476  & 6.89  & 5.10  & 0.74  & 3.31 & N  \\
2 & 16 26 10.3  & -24 23 12.0  & 3.41  & 0.87  & 99.7  & 0.0604  & 
 1.19  & 0.764  & 7.07  & 11.6  & 1.64  & 1.33 & Y  \\
3 & 16 26 13.3  & -24 20 6.7  & 3.02  & 0.70  & 72.7  & 0.0441  & 
1.67  & 0.502  & 2.78  & 5.32  & 1.91  & 1.34  & N  \\
4 & 16 26 14.8  & -24 21 8.3  & 3.41  & 0.78  & 78.8  & 0.0478  & 
1.05  & 0.551  & 3.85  & 6.28  & 1.63  & 1.47  & N  \\
5 & 16 26 19.4  & -24 23 32.0  & 2.76  & 0.69  & 90.0  & 0.0545  & 
1.52  & 0.528  & 4.57  & 6.89  & 1.51  & 1.17  & Y \\
6 & 16 26 22.4  & -24 24 54.0  & 3.28  & 1.05  & 76.8  & 0.0466  & 
1.34  & 0.730  & 4.68  & 8.35  & 1.79  & 1.92 &  Y \\
7 & 16 26 23.8  & -24 20 26.5  & 3.41  & 0.99  & 73.3  & 0.0444  & 
1.07  & 0.639  & 4.06  & 6.82  & 1.68  & 1.92  & N  \\
8 & 16 26 26.7  & -24 16 40.1  & 3.80  & 0.53  & 36.5  & 0.0221  & 
1.42  & 0.453  & 0.49  & 2.46  & 5.03  & 1.86  & N \\
9 & 16 26 26.9  & -24 22 29.8  & 3.28  & 3.27  & 83.2  & 0.0504  & 
1.34  & 0.547  & 17.88  & 6.59  & 0.37  & 5.77  & Y \\
10 & 16 26 26.9  & -24 23 52.0  & 3.80  & 3.99  & 90.0  & 0.0545  & 
1.20  & 0.627  & 21.78  & 8.19  & 0.38  & 5.56  & Y \\
11 & 16 26 28.4  & -24 23 31.4  & 2.76  & 0.79  & 65.9  & 0.0399  & 
1.62  & 0.585  & 1.97  & 5.58  & 2.83  & 1.28  & Y  \\
12 & 16 26 33.0  & -24 24 12.2  & 3.02  & 0.99  & 67.6  & 0.0410  & 
1.28  & 0.415  & 2.43  & 4.27  & 1.76  & 1.46  & N  \\
13 & 16 26 33.0  & -24 26 15.6  & 3.15  & 1.12  & 61.3  & 0.0371  & 
1.26  & 0.609  & 2.73  & 5.41  & 1.98  & 2.21  & N  \\
14 & 16 26 34.5  & -24 25 55.0  & 3.54  & 1.39  & 86.3  & 0.0523  & 
1.11  & 0.442  & 6.25  & 5.70  & 0.91  & 1.81  & N  \\
15 & 16 26 36.0  & -24 25 34.3  & 3.15  & 0.69  & 51.2  & 0.0311  & 
1.23  & 0.267  & 1.03  & 2.58  & 2.52  & 1.42  & N  \\
16 & 16 26 43.5  & -24 25 54.3  & 3.80  & 0.78  & 56.1  & 0.0340  & 
1.19  & 0.279  & 1.38  & 2.87  & 2.08  & 1.46  & N  \\
17 & 16 26 45.1  & -24 26 56.0  & 3.93  & 0.83  & 37.4  & 0.0227  & 
1.96  & 0.186  & 0.72  & 1.71  & 2.39  & 2.55  & N \\
18 & 16 26 45.3  & -24 33 26.8  & 3.80  & 1.17  & 113.8  & 0.0690  & 
1.24  & 0.528  & 7.02  & 7.15  & 1.02  & 0.89  & Y  \\
19 & 16 26 46.5  & -24 23 50.7  & 3.02  & 0.63  & 50.1  & 0.0304  & 
2.28  & 0.388  & 0.93  & 3.02  & 3.24  & 1.38  & N\\
20 & 16 26 46.6  & -24 27 57.6  & 3.80  & 0.53  & 50.3  & 0.0305  & 
1.27  & 0.316  & 0.39  & 2.02  & 5.13  & 0.58  & Y \\
21 & 16 26 49.6  & -24 26 55.7  & 3.93  & 0.85  & 43.9  & 0.0266  & 
1.11  & 0.454  & 0.79  & 2.36  & 2.98  & 1.73  & N \\
22 & 16 26 49.7  & -24 30 0.8  & 3.80  & 1.23  & 99.4  & 0.0602  & 
1.03  & 0.385  & 4.14  & 4.60  & 1.11  & 0.79  & Y \\
23 & 16 26 52.9  & -24 36 52.0  & 4.45  & 0.92  & 58.2  & 0.0353  & 
1.37  & 0.550  & 0.49  & 3.83  & 7.76  & 0.47  & N \\
24 & 16 26 52.9  & -24 37 12.5  & 4.71  & 0.98  & 50.7  & 0.0307  & 
2.10  & 0.337  & 0.77  & 2.13  & 2.76  & 1.10  & N \\
25 & 16 26 55.9  & -24 36 31.2  & 4.58  & 0.95  & 63.6  & 0.0386  & 
1.37  & 0.438  & 1.03  & 3.30  & 3.20  & 0.75  & N \\
26 & 16 26 57.0  & -24 23 50.0  & 3.80  & 1.37  & 79.1  & 0.0480  & 
1.02  & 0.393  & 2.41  & 3.72  & 1.54  & 0.91  & Y \\
27 & 16 26 57.3  & -24 31 22.6  & 3.93  & 1.52  & 101.9  & 0.0618  & 
1.21  & 0.543  & 7.63  & 6.61  & 0.87  & 1.34  & Y \\
28 & 16 26 57.4  & -24 37 53.4  & 4.58  & 0.66  & 48.6  & 0.0295  & 
1.06  & 0.666  & 0.74  & 4.07  & 5.54  & 1.19  & Y \\
29 & 16 27 1.9  & -24 34 47.9  & 3.67  & 1.56  & 100.5  & 0.0609  & 
1.24  & 0.669  & 6.09  & 8.46  & 1.39  & 1.11  & Y \\
30 & 16 27 3.4  & -24 36 10.1  & 4.97  & 0.95  & 58.1  & 0.0352  & 
1.28  & 0.379  & 1.19  & 2.66  & 2.23  & 1.13  & Y \\
31 & 16 27 3.5  & -24 38 54.7  & 4.58  & 1.29  & 56.5  & 0.0343  & 
1.65  & 0.427  & 1.89  & 2.87  & 1.51  & 1.95  & N \\
32 & 16 27 4.7  & -24 26 54.7  & 3.67  & 0.47  & 72.2  & 0.0438  & 
1.07  & 0.547  & 0.94  & 4.73  & 5.05  & 0.46  & Y \\
33 & 16 27 6.3  & -24 32 44.2  & 4.06  & 1.52  & 87.0  & 0.0528  & 
1.14  & 0.523  & 4.53  & 5.41  & 1.20  & 1.28  & N \\
34 & 16 27 9.6  & -24 39 35.4  & 4.32  & 1.03  & 78.5  & 0.0476  & 
1.11  & 0.587  & 1.69  & 5.59  & 3.31  & 0.65  & Y \\
35 & 16 27 10.7  & -24 27 35.4  & 4.32  & 0.71  & 81.4  & 0.0493  & 
1.29  & 0.516  & 1.85  & 4.99  & 2.70  & 0.64  & Y \\
36 & 16 27 10.8  & -24 29 18.2  & 3.80  & 1.38  & 73.9  & 0.0448  & 
1.19  & 0.765  & 3.92  & 7.51  & 1.92  & 1.81  & Y \\
37 & 16 27 12.2  & -24 25 52.4  & 3.41  & 0.77  & 78.8  & 0.0478  & 
1.15  & 0.472  & 1.18  & 4.39  & 3.72  & 0.45  & Y \\
38 & 16 27 12.3  & -24 29 59.3  & 3.80  & 1.37  & 69.2  & 0.0420  & 
1.49  & 0.722  & 3.17  & 6.48  & 2.04 & 1.78  & N \\
39 & 16 27 14.0  & -24 38 12.8  & 4.71  & 0.65  & 53.9  & 0.0327  & 
2.27  & 0.616  & 0.69  & 4.07  & 5.94  & 0.81  & N \\
40 & 16 27 15.3  & -24 30 40.2  & 3.67  & 1.37  & 100.0  & 0.0606  & 
1.21  & 0.678  & 5.61  & 8.58  & 1.53  & 1.05  & Y\\ 
41 & 16 27 18.2  & -24 27 55.5  & 3.41  & 0.78  & 92.1  & 0.0558  & 
1.13  & 0.723  & 2.35  & 8.63  & 3.67  & 0.56  & Y \\
42 & 16 27 21.1  & -24 23 48.4  & 3.15  & 1.15  & 79.3  & 0.0481  & 
1.08  & 0.471  & 2.87  & 4.41  & 1.54  & 1.07  & Y \\
43 & 16 27 21.2  & -24 26 53.6  & 4.45  & 1.26  & 98.0  & 0.0594  & 
1.15  & 0.596  & 5.02  & 7.11  & 1.42  & 0.99  & Y \\
44 & 16 27 21.6  & -24 39 55.2  & 4.84  & 0.96  & 70.6  & 0.0428  & 
1.11  & 0.735  & 1.66  & 6.78  & 4.07  & 0.88  & Y \\
45 & 16 27 24.3  & -24 27 55.1  & 3.80  & 1.76  & 82.2  & 0.0498  & 
1.04  & 0.527  & 4.33  & 5.15  & 1.19  & 1.45  & Y \\
46 & 16 27 24.7  & -24 40 56.7  & 4.84  & 0.83  & 58.3  & 0.0353  & 
1.26  & 0.631  & 0.95  & 4.55  & 4.81  & 0.89  & Y \\
47 & 16 27 25.7  & -24 24 29.3  & 3.02  & 1.70  & 95.7  & 0.0580  & 
1.15  & 0.438  & 5.37  & 4.96  & 0.92  & 1.14  & Y \\
48 & 16 27 25.8  & -24 27 13.8  & 3.80  & 1.93  & 71.7  & 0.0435  & 
1.12  & 0.595  & 5.47  & 5.19  & 0.95  & 2.76  & Y \\
49 & 16 27 26.0  & -24 33 24.1  & 3.54  & 0.97  & 100.9  & 0.0612  & 
1.56  & 0.449  & 3.02  & 5.35  & 1.77  & 0.55  & N \\
50 & 16 27 28.8  & -24 27 34.2  & 4.45  & 1.87  & 86.9  & 0.0527  & 
1.21  & 0.405  & 4.58  & 4.20  & 0.92  & 1.30  & Y \\
51 & 16 27 29.2  & -24 41 37.5  & 3.67  & 0.79  & 86.1  & 0.0522  & 
1.13  & 0.525  & 2.27  & 5.37  & 2.37  & 0.66 & Y \\
52 & 16 27 30.7  & -24 40 35.7  & 4.32  & 1.38  & 84.0  & 0.0509  & 
1.62  & 0.316  & 2.53  & 3.37  & 1.33  & 0.79  & N \\
53 & 16 27 33.3  & -24 26 11.6  & 3.93  & 1.63  & 114.2  & 0.0692  & 
1.46  & 0.563  & 8.99  & 7.72  & 0.86  & 1.12  & Y \\
54 & 16 27 33.3  & -24 26 52.7  & 4.58  & 1.97  & 95.6  & 0.0579  & 
1.01  & 0.361  & 4.96  & 4.21  & 0.85  & 1.06  & N  \\
55 & 16 27 33.3  & -24 28 35.6  & 3.80  & 0.76  & 92.1  & 0.0558  & 
1.12  & 0.675  & 2.01  & 7.85  & 3.90  & 0.48  & N \\
56 & 16 27 36.5  & -24 34 4.5  & 3.41  & 0.51  & 76.6  & 0.0465  & 
2.33  & 0.603  & 0.90  & 5.64  & 6.29  & 0.37  & N \\
57 & 16 27 38.3  & -24 41 16.4  & 3.54  & 0.53  & 57.7  & 0.0350  & 
1.61  & 0.331  & 0.47  & 2.39  & 5.10  & 0.45  & N \\
58 & 16 27 42.9  & -24 42 38.3  & 3.80  & 1.06  & 76.8  & 0.0466  & 
1.51 & 0.444  & 3.19  & 4.03  & 1.26  & 1.31 & Y \\
59 & 16 27 46.0  & -24 44 41.6  & 3.67  & 1.31  & 80.8  & 0.0490  & 
1.53  & 0.302  & 2.97  & 3.16  & 1.06  & 1.05  & Y \\
60& 16 27 51.9  & -24 42 58.3  & 3.93  & 1.02  & 75.1  & 0.0455  & 
1.02  & 0.341  & 1.70  & 3.17  & 1.87  & 0.75  & N \\
61& 16 27 57.7  & -24 35 4.8  & 4.19  & 0.97  & 56.9  & 0.0345  & 
1.44  & 0.398  & 1.31  & 2.71  & 2.06  & 1.32  & N \\
62& 16 27 59.2  & -24 34 23.6  & 4.19  & 0.75  & 41.0  & 0.0248  & 
1.01  & 0.353  & 0.64  & 1.78  & 2.80  & 1.72  & N \\
63& 16 28 0.6 & -24 33 1.2 & 4.32  & 1.34  & 73.0  & 0.0442  & 
1.05  & 0.402  & 2.43  & 3.51  & 1.44  & 1.16 & N  \\
64& 16 28 3.7  & -24 34 43.8  & 4.19  & 0.75  & 48.5  & 0.0294  & 
1.68  & 0.238  & 0.60  & 1.68  & 2.79  & 0.99  & N \\
65& 16 28 6.7  & -24 34 23.1  & 4.32  & 0.71  & 62.9  & 0.0381  & 
1.14  & 0.350  & 0.91  & 2.71  & 2.97  & 0.68  & N \\
66& 16 28 15.8  & -24 35 44.7  & 4.19  & 1.06  & 69.5  & 0.0421  & 
1.04  & 0.306  & 1.38  & 2.74  & 1.98  & 0.77  & N \\
67& 16 28 20.4  & -24 36 25.6  & 4.19  & 1.60  & 52.6  & 0.0319  & 
1.11  & 0.324  & 1.94  & 2.15  & 1.11 & 2.48  & Y \\
68& 16 28 26.4  & -24 36 4.6  & 4.19  & 1.20  & 97.4  & 0.0590  & 
1.34  & 0.271  & 3.42  & 3.58  & 1.05  & 0.69  & N \\
\enddata
\tablecomments{Units of right ascension are hours, minutes, and seconds, 
and units of declination are degrees, arcminutes, and arcseconds.
Cores No. 1 through 17 and 19 are located in the Oph A region and 
the temperature is set to be 18K. For the other cores, $T=12$K.
The typical uncertainties of $R_{\rm core}$ and $dv_{\rm core}$
are, respectively, about 0.01 pc, derived from the uncertainty in the
 estimation of the core projected area, and the velocity resolution of  
0.13 km s$^{-1}$.}
\tablenotetext{a}{Y and N mean, respectively, the cores associated 
with and without YSOs that are classified as either Class I or Flat 
Spectrum objects based on the Spitzer observations.
}
\end{deluxetable}

\begin{deluxetable}{lllll}
%\tabletypesize{\scriptsize}
%\rotate
\tablecolumns{5}
\tablecaption{Summary of the Physical Properties 
of the H$^{13}$CO$^+$ cores\label{tab:h13co+core}}
\tablewidth{\columnwidth}
\tablehead{\colhead{Property} & \colhead{Minimum} & \colhead{Maximum}
 &\colhead{Mean\tablenotemark{a}}   &\colhead{Median} 
}
\startdata
$R_{\rm core}$ (pc) & 0.0221 & 0.0692  & 0.0450 $\pm$ 0.0113  & 0.0451 \\
$dv_{\rm core}$ (km s$^{-1}$) & 0.186 & 0.765  & 0.488 $\pm$ 0.142  & 0.474 \\
$M_{\rm LTE}$ (M$_\odot$) & 0.39 & 21.79  & 3.35 $\pm$ 3.57  & 2.42 \\
$M_{\rm vir}/M_{\rm LTE}$  & 0.368 & 7.76  & 2.37 $\pm$ 1.57 & 1.89 \\
$\bar{n}$ (cm$^{-3}$) & $3.70 \times 10^4$ & $5.77\times 10^5$ & 
$(1.35\pm 0.96) \times 10^5$ & $1.35\times 10^5$ \\
Aspect Ratio & 1.01 & 2.33  & 1.34$\pm$ 0.33  & 1.22 \\
\enddata
\tablenotetext{a}{With standard deviation}
%\tablecomments{..........}
\end{deluxetable}

\begin{deluxetable}{lllll}
%\tabletypesize{\scriptsize}
%\rotate
\tablecolumns{5}
\tablecaption{Summary of the Physical Properties 
of the 850$\mu m$ cores\label{tab:scubacore}}
\tablewidth{\columnwidth}
\tablehead{\colhead{Property} & \colhead{Minimum} & \colhead{Maximum}
 &\colhead{Mean\tablenotemark{a}}   &\colhead{Median} 
}
\startdata
$R_{\rm core}$ (pc) & 0.00494 & 0.0601  & 0.0204 $\pm$ 0.0118  & 0.0174 \\
$M_{\rm LTE}$ (M$_\odot$) & 0.0868 & 8.25  & 1.42 $\pm$ 2.13  & 0.553 \\
$\bar{n}$ (cm$^{-3}$) & $1.41 \times 10^5$ & $3.20\times 10^6$ & 
$(6.53\pm 6.29) \times 10^5$ & $4.37\times 10^5$ \\
\enddata
\tablenotetext{a}{With standard deviation}
\tablecomments{The uniform temperature of $T=15$ K is assumed
for all the cores in \citet{jorgensen08}.
}
\end{deluxetable}

\begin{deluxetable}{lllll}
%\tabletypesize{\scriptsize}
%\rotate
\tablecolumns{5}
\tablecaption{Summary of the Physical Properties 
of the H$^{13}$CO$^+$ cores with and without YSOs\label{tab:yso}}
\tablewidth{\columnwidth}
\tablehead{\colhead{Property} & \colhead{Minimum} & \colhead{Maximum}
 &\colhead{Mean\tablenotemark{a}}   &\colhead{Median} 
%\colhead{Starless} & \colhead{} & \colhead{}
% &\colhead{}   &\colhead{} \\
}
\startdata
Cores without YSOs & & & & \\
\hline
$R_{\rm core}$ (pc) & 0.0221 & 0.0612  & 0.0403 $\pm$ 0.0105  & 0.0398 \\
$dv_{\rm core}$ (km s$^{-1}$) & 0.186 & 0.722  & 0.431 $\pm$ 0.132 &
 0.421 \\
$M_{\rm LTE}$ (M$_\odot$) & 0.469 & 6.89 & 1.54 $\pm$ 1.69  & 1.54 \\
$M_{\rm vir}/M_{\rm LTE}$ & 0.74 & 7.76  & 2.63 $\pm$ 1.65   & 2.05 \\
$\bar{n}$ (cm$^{-3}$) & $3.70 \times 10^4$ & $3.31\times 10^5$ & 
$(1.29\pm 0.66) \times 10^5$ & $1.30\times 10^5$ \\
Aspect Ratio & 1.01 & 2.33  & 1.45$\pm$ 0.417  & 1.36 \\[1mm]
\hline
Cores with YSOs& & & &  \\[1mm]
\hline
$R_{\rm core}$ (pc)& 0.0295 & 0.0692  & 0.0497 $\pm$ 0.0102  & 0.0492 \\
$dv_{\rm core}$ (km s$^{-1}$)& 0.302 & 0.765  & 0.544 $\pm$ 0.130  & 0.545 \\
$M_{\rm LTE}$ (M$_\odot$) & 0.394 & 21.8  & 4.55 $\pm$ 4.48  & 3.55 \\
$M_{\rm vir}/M_{\rm LTE}$ & 0.37 & 5.54 & 2.11 $\pm$ 1.46 & 1.53 \\
$\bar{n}$ (cm$^{-3}$) & $0.449 \times 10^5$ & $5.77\times 10^5$ & 
$(1.40\pm 1.20) \times 10^5$ & $1.12\times 10^5$ \\
Aspect Ratio & 1.02 & 1.62 & 1.22$\pm$ 0.155  & 1.19
\enddata
\tablenotetext{a}{With standard deviation}
%\tablecomments{..........}
\end{deluxetable}

\end{document}